\documentclass[prd,showpacs,floatfix,twocolumn,,amsmath,amssymb,floatfix]{revtex4}
\usepackage{graphicx,color,dcolumn,booktabs,bm}
\usepackage{longtable,lscape}
\usepackage{txfonts}
\usepackage{overpic}
\usepackage{amssymb}
\usepackage{indentfirst}
\usepackage{epsfig}
\usepackage{epstopdf}   %{feynmp}
\usepackage{slashed}  %for Feynman symbols
\usepackage{cases}
\usepackage{color}
\usepackage{multirow}
\usepackage{graphicx,color,dcolumn,booktabs,bm}

\graphicspath{{Figures/}} %
\usepackage[colorlinks, citecolor=blue,anchorcolor=red,menucolor=red, linkcolor=red,filecolor=red,runcolor=red,urlcolor=blue,frenchlinks=red]{hyperref}
\newcommand{\minitab}[2][l]{\begin{tabular}{#1}#2\end{tabular}}
\begin{document}

\title{
The possible members of the $5^1S_0$ meson nonet}

\author{Shi-Chen Xue}
\affiliation{School of Physics and Engineering, Zhengzhou University, Zhengzhou, Henan 450001, China}

\author{Guan-Ying Wang}
\affiliation{School of Physics and Engineering, Zhengzhou University, Zhengzhou, Henan 450001, China}

\author{Guan-Nan Li}
\affiliation{School of Physics and Engineering, Zhengzhou University, Zhengzhou, Henan 450001, China}

\author{En Wang}\email[]{Corresponding Author: wangen@zzu.edu.cn}
\affiliation{School of Physics and Engineering, Zhengzhou University, Zhengzhou, Henan 450001, China}

\author{De-Min Li} \email[]{Corresponding Author: lidm@zzu.edu.cn}
\affiliation{School of Physics and Engineering, Zhengzhou University, Zhengzhou, Henan 450001, China}

\date{\today}

\begin{abstract}
The strong decays of the $5^1S_0$ $q\bar{q}$ states are evaluated in the $^3P_0$ model with two types of space wave functions. Comparing the model expectations with the experimental data for the $\pi(2360)$, $\eta(2320)$, $X(2370)$, and $X(2500)$, we suggest
that the $\pi(2360)$, $\eta(2320)$, and $X(2500)$ can be assigned as the members of the $5^1S_0$ meson nonet, while the $5^1S_0$ assignment for the $X(2370)$ is not favored by its width. The $5^1S_0$ kaon is predicted to have a mass of about 2418 MeV and a width of about 163 MeV or 225 MeV.
\end{abstract}
\maketitle

\section{Introduction}{\label{Introduction}}

In the framework of quantum chromodynamics (QCD), apart from the ordinary $q\bar{q}$ states,
other exotic states such as glueballs, hybrids, and tetraquarks are permitted
to exist in meson spectra. To identify these exotic states, one needs to distinguish them from the
background of ordinary $q\bar{q}$ states, which requires one to understand well the conventional $q\bar{q}$ meson
spectroscopy both theoretically and experimentally. To be able to understand the nature of a newly observed state, it is natural and necessary to exhaust the possible $q\bar{q}$ description before restoring to more exotic assignments.

\begin{table}[tpbh]
\begin{center}
\caption{ \label{tab:pseu}The pseudoscalar states reported experimentally.}
%\footnotesize
\begin{tabular}{c|l} \hline\hline Isospin & \hspace*{3cm}~States  \\\hline
$I=1$  & $\pi$, $\pi(1300)$, $\pi(1800)$, $\pi(2070)$, $\pi(2360)$ \\\hline
\multirow{2}*{\minitab[c]{$I=0$}}   & $\eta$, $\eta(1295)$, $\eta(1760)$, $\eta(2010)$,
 $\eta(2100)$, $\eta(2190)$, $\eta(2320)$ \\
 & $\eta'$, $\eta(1475)$, $X(1835)$\footnote{ The spin-parity of the $X(1835)$ is not determined experimentally,
but the angular distribution of the radiative
photon is consistent with expectations for a pseudoscalar~\cite{Ablikim:2010au}.},  $\eta(2225)$, $X(2500)$ \\\hline
 $I=1/2$& $K$, $K(1460)$, $K(1830)$ \\
 \hline\hline
\end{tabular}
\end{center}
\end{table}

 As shown in Table~\ref{tab:pseu}, many pseudoscalar states have been accumulated experimentally~\cite{PDG2016,Ablikim:2016hlu}. Among these states,
 the assignments of the $\pi$, $\eta$, $\eta'$, and $K$ as the members of the $1^1S_0$
 meson nonet and the $\pi(1300)$, $\eta(1295)$, $\eta(1475)$, and $K(1460)$  as the
 members of the $2^1S_0$ meson nonet have been widely accepted~\cite{PDG2016}. In our previous works, we
 suggested that the $\pi(1800)$, $\eta(1760)$, $X(1835)$, and $K(1830)$ can be identified as
 members of the $3^1S_0$ meson nonet~\cite{Li:2008mza}, the $\pi(2070)$, $\eta(2100)$, and $\eta(2225)$
 can be identified as the members of $4^1S_0$ meson~\cite{Li:2008et}, and the $X(2500)$
 is the mainly $s\bar{s}$ member of the $5^1S_0$ meson nonet~\cite{Pan:2016bac}, where the mixing of the $X(2500)$ and its isoscalar partner is not considered and other members of the $5^1S_0$ meson nonet are not discussed. In this work, we shall address the possible SU(3) multiplet partners of the $X(2500)$. With the assignment of the $X(2500)$ as the $s\bar{s}$ member of the $5^1S_0$ nonet, one can expect that other members of the $5^1S_0$ nonet should be lighter than
 the $X(2500)$. Along this line, considering that other pseudoscalar states have discussed in our previous works~\cite{Li:2008mza,Li:2008et}, we shall focus on the $\pi(2360)$ and $\eta(2320)$ shown in Table~\ref{tab:pseu}, and check whether they can be explained as the $5^1S_0$ $q\bar{q}$ states or not. Study on the pseudoscalar radial $q\bar{q}$ excitations in the mass region of $2.3\sim 2.6 $ GeV is especially interesting because the pseudoscalar glueball is predicted to exist in this mass region~\cite{Morningstar:1999rf,Hart:2001fp,Chen:2005mg}.

The $\pi(2360)$ was observed in $\bar{p}p \to 3\pi^0,\pi^0\eta, \pi^0\eta^\prime, \eta \eta \pi^0$, and its mass and width are $2360\pm 25$ MeV and $300^{+100}_{-50}$ MeV, respectively~\cite{Anisovich:2001pn,Anisovich:2001pp}. The $\pi(5^1S_0)$ mass is expected to be $2316 \pm 40$~MeV in a relativistic independent quark model~\cite{Khruschov:2005zc} or $2385$ MeV in a relativistic quark model~\cite{Ebert:2009ub}, both consistent with the $\pi(2360)$ mass. Thus, the $\pi(2360)$ appears a good candidate for the $\pi(5^1S_0)$ based on its measured mass.

\begin{figure}[htb]
\centering
\includegraphics[scale=0.35]{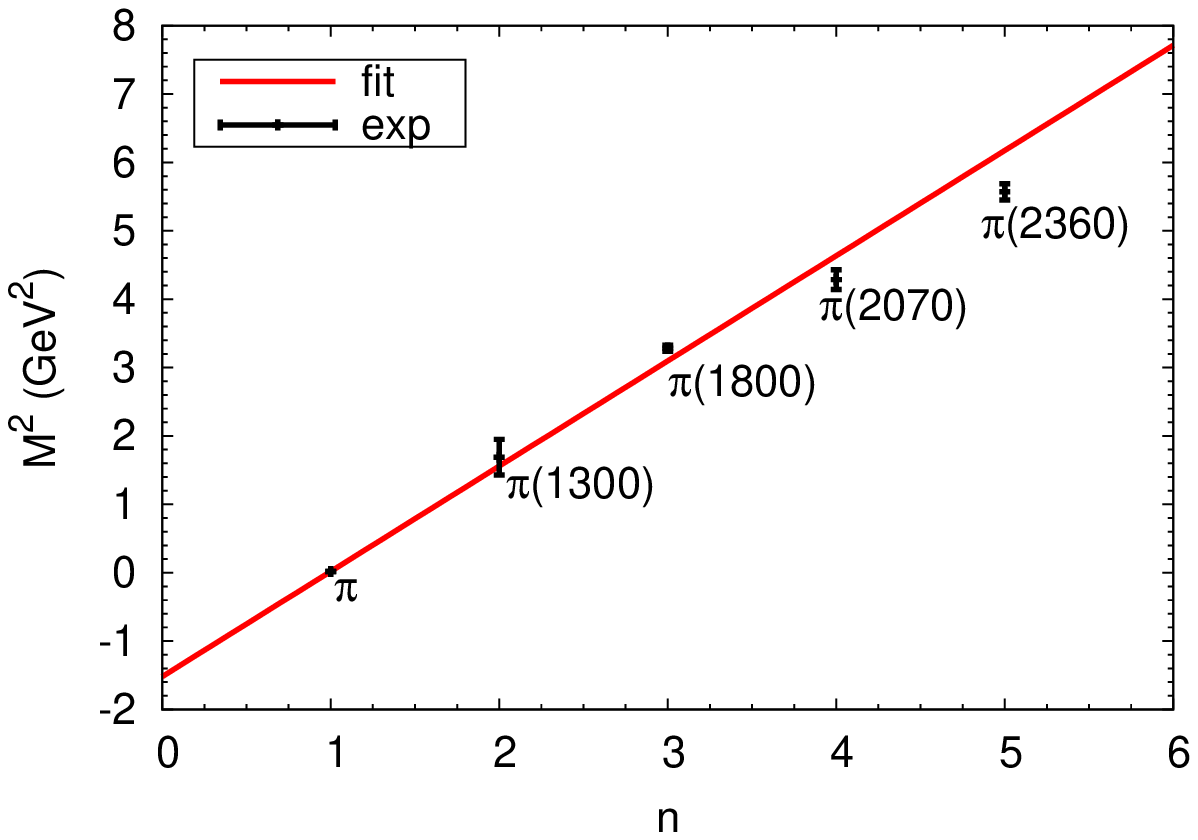}\includegraphics[scale=0.35]{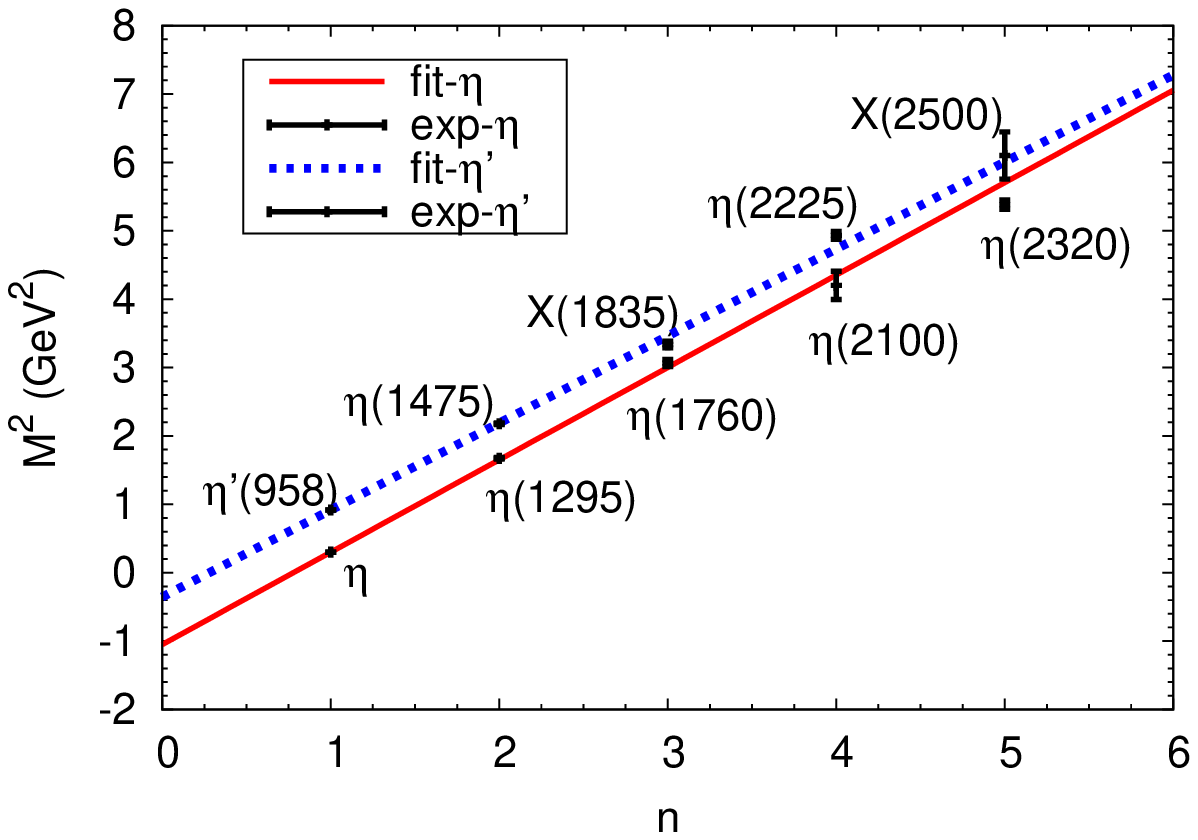}
\caption{The $\pi$, $\eta$, and $\eta^\prime$-trajectories with $M^2_n = M^2_0+(n-1)\mu^2$ by fitting to the experimental masses of the mesons. $\pi$-trajectory: $M^2_0=0.019480\pm0.000001~\mbox{GeV}^2$, $\mu^2=1.5387\pm 0.0165~\mbox{GeV}^2$, $\chi^2/\mbox{d.o.f}=49.7/(5-2)$; $\eta$-trajectory: $M^2_0=0.30015\pm0.00002~\mbox{GeV}^2$, $\mu^2=1.3511\pm 0.0084 ~\mbox{GeV}^2$, $\chi^2/\mbox{d.o.f}=27.9/(5-2)$; $\eta^\prime$-trajectory: $M^2_0=0.91734\pm0.00115~\mbox{ GeV}^2$, $\mu^2=1.2723\pm 0.0092~\mbox{GeV}^2$, $\chi^2/\mbox{d.o.f}=19.9/(5-2)$. The meson masses used to fit are taken from Refs.~\cite{PDG2016,Ablikim:2016hlu}.}
\label{fig:piRegge}
\end{figure}

A series of the papers of Anisovich~\cite{Anisovich:2000kxa,Anisovich:2001ig,Anisovich:2002us,Anisovich:2003tm,Anisovich:2004vj} indicate that with a good accuracy, the light $q\bar{q}$ meson states with different
radial excitations fit to the following quasi-linear $(n,M^2_n)$-trajectories
\begin{equation}
M^2_n=M^2_0+(n-1)\mu^2,  \label{eq:regge}
\end{equation}
where $M_n$ denotes the mass of the meson with radial quantum number $n$, $M^2_0$ and $\mu^2$ are the parameters of the corresponding trajectory. The relation of Eq.~(\ref{eq:regge}) can be derived from the Regge phenomenology~\cite{collins,Li:2007px}. One can use this relation to roughly estimate the masses for higher radial excitations. As displayed in Fig.~\ref{fig:piRegge}, we find that in the $(n,M^2_n)$ plane,
the three pseudoscalar meson groups, [$\pi$, $\pi(1300)$, $\pi(1800)$, $\pi(2070)$, $\pi(2360)$],
[$\eta$, $\eta(1295)$, $\eta(1760)$, $\eta(2100)$, $\eta(2320)$],
and [$\eta^\prime$, $\eta(1475)$, $X(1835)$, $\eta(2225)$, $X(2500)$], approximately populate the $\pi$, $\eta$, and $\eta^\prime$-linear trajectories, respectively. With the assignment that the [$\pi(1800)$, $\eta(1760)$, $X(1835)$] and [$\pi(2070)$, $\eta(2100)$, $\eta(2225)$ ] belong to the $3^1S_0$ and $4^1S_0$ meson nonets, respectively, one can naturally expect that the $\pi(2360)$, $\eta(2320)$, and $X(2500)$ could belong to the $5^1S_0$ nonet based on their masses.

Both the mass and width of a resonance are related to its inner structure. Although the masses of the $\pi(2360)$, $\eta(2320)$, and $X(2500)$ are consistent with them belonging to the $5^1S_0$ meson nonet, their decay properties also need to be compared with model expectations in order to identify the possible candidates for the $5^1S_0$ meson nonet. Below, we shall evaluate their strong decays in the framework of the $^3P_0$ model.

This paper is organized as follows. In Sec.~\ref{sec:formalisms}, we present the $^3P_0$ model parameters used in our calculations. The results and discussions are given in Sec.~\ref{sec:result}. Finally, a short summary is given in Sec.~\ref{sec:summary}.

\section{Model and Parameters}
\label{sec:formalisms}

The $^3P_0$ model has been widely used to study the strong decays of hadrons in literature~\cite{Roberts:1992js,Blundell:1996as,Barnes:1996ff,Barnes:2002mu,
Close:2005se,Barnes:2005pb,Zhang:2006yj,Ding:2007pc,Li:2008mza,Li:2008we,Li:2008et,Li:2008xy,
Li:2009rka,Li:2009qu,Li:2010vx,Lu:2014zua,Pan:2016bac,Lu:2016bbk,Wang:2017pxm}. In the $^3P_0$ model, the meson strong decay takes place by producing a quark-antiquark pair with vacuum quantum number $J^{PC}=0^{++}$. The newly produced quark-antiquark pair, together with the $q\bar{q}$ within the initial meson, regroups into two outgoing mesons in all possible quark rearrangement ways. Some detailed reviews on the $^3P_0$ model can be found in Refs.~\cite{Li:2008mza,Li:2008et,Li:2008we,Roberts:1992js,Blundell:1996as}. Here we give the main ingredients of the $^3P_0$ model briefly.

Following the conventions in Ref.~\cite{Li:2008mza}, the transition operator $T$ of the decay  $A\rightarrow BC$ in the $^3P_0$ model is given by
\begin{eqnarray}
T=-3\gamma\sum_m\langle 1m1-m|00\rangle\int
d^3\boldsymbol{p}_3d^3\boldsymbol{p}_4\delta^3(\boldsymbol{p}_3+\boldsymbol{p}_4)\nonumber\\
{\cal{Y}}^m_1\left(\frac{\boldsymbol{p}_3-\boldsymbol{p}_4}{2}\right
)\chi^{34}_{1-m}\phi^{34}_0\omega^{34}_0b^\dagger_3(\boldsymbol{p}_3)d^\dagger_4(\boldsymbol{p}_4),
\end{eqnarray}
where the $\gamma$ is a dimensionless parameter denoting the production strength of the quark-antiquark pair $q_3\bar{q}_4$ with quantum number $J^{PC}=0^{++}$. $\boldsymbol{p}_3$ and  $\boldsymbol{p}_4$ are the momenta of the created quark  $q_3$ and  antiquark $\bar{q}_4$, respectively. $\chi^{34}_{1,-m}$, $\phi^{34}_0$, and $\omega^{34}_0$ are the spin, flavor, and color wave functions of $q_3\bar{q}_4$, respectively. The solid harmonic polynomial  ${\cal{Y}}^m_1(\boldsymbol{p})\equiv|\boldsymbol{p}|^1Y^m_1(\theta_p, \phi_p)$ reflects the momentum-space distribution of the $q_3\bar{q_4}$.

The $S$ matrix of the process $A\rightarrow BC$ is defined by
\begin{eqnarray}
\langle BC|S|A\rangle=I-2\pi i\delta(E_A-E_B-E_C)\langle BC|T|A\rangle,
\end{eqnarray}
where $|A\rangle$ ($|B\rangle$,$|C\rangle$) is the mock meson defined by ~\cite{Hayne:1981zy}
\begin{eqnarray}
&&|A(n^{2S_A+1}_AL_{A}\,\mbox{}_{J_A M_{J_A}})(\boldsymbol{P}_A)\rangle
\equiv \nonumber\\
&& \sqrt{2E_A}\sum_{M_{L_A},M_{S_A}}\langle L_A M_{L_A} S_A
M_{S_A}|J_A
M_{J_A}\rangle\nonumber\\
&&\times  \int d^3\boldsymbol{p}_A\psi_{n_AL_AM_{L_A}}(\boldsymbol{p}_A)\chi^{12}_{S_AM_{S_A}}
\phi^{12}_A\omega^{12}_A\nonumber\\
&&\times  \left|q_1\left({\scriptstyle
\frac{m_1}{m_1+m_2}}\boldsymbol{P}_A+\boldsymbol{p}_A\right)\bar{q}_2
\left({\scriptstyle\frac{m_2}{m_1+m_2}}\boldsymbol{P}_A-\boldsymbol{p}_A\right)\right\rangle.
\end{eqnarray}
Here $m_1$ and $m_2$ ($\boldsymbol{p}_1$ and $\boldsymbol{p}_2$) are the masses (momenta) of the
quark $q_1$ and the antiquark $\bar{q}_2$, respectively; $\boldsymbol{P}_A=\boldsymbol{p}_1+\boldsymbol{p}_2$,
$\boldsymbol{p}_A=\frac{m_2\boldsymbol{p}_1-m_1\boldsymbol{p}_2}{m_1+m_2}$;
$\chi^{12}_{S_AM_{S_A}}$, $\phi^{12}_A$, $\omega^{12}_A$, and
$\psi_{n_AL_AM_{L_A}}(\boldsymbol{p}_A)$ are the spin, flavor, color, and
space wave functions of the meson $A$ composed of $q_1\bar{q}_2$ with total energy $E_A$, respectively. $n_A$ is the radial quantum number of the meson $A$. $\boldsymbol{S}_A=\boldsymbol{s}_{q_1}+\boldsymbol{s}_{\bar{q}_2}$, $\boldsymbol{J}_A=\boldsymbol{L}_A+\boldsymbol{S}_A$, $\boldsymbol{s}_{q_1}(\boldsymbol{s}_{\bar{q}_2})$ is the spin of $q_1(\bar{q}_2)$, and $\boldsymbol{L}_A$ is the relative orbital angular momentum between $q_1$ and $\bar{q}_2$. $\langle L_A M_{L_A} S_AM_{S_A}|J_AM_{J_A}\rangle$ denotes a Clebsch-Gordan coefficient.

The transition matrix element $\langle BC|T|A\rangle$ can be written as
\begin{eqnarray}
\langle BC|T|A\rangle=\delta^3(\boldsymbol{P}_A-\boldsymbol{P}_B-\boldsymbol{P}_C){\cal{M}}^{M_{J_A}M_{J_B}M_{J_C}}(\boldsymbol{P}),
\end{eqnarray}
where ${\cal{M}}^{M_{J_A}M_{J_B}M_{J_C}}
(\boldsymbol{P})$ is the helicity amplitude. In the center of mass frame of meson A, the helicity amplitude is
\begin{eqnarray}
&&{\cal{M}}^{M_{J_A}M_{J_B}M_{J_C}}(\boldsymbol{P})=\gamma\sqrt{8E_AE_BE_C} \sum_{M_{L_A},M_{S_A}}
\nonumber\\&&\times \sum_{M_{L_B},M_{S_B}} \sum_{M_{L_C},M_{S_C}}\sum_m
\langle L_A M_{L_A} S_AM_{S_A}|J_AM_{J_A}\rangle\nonumber\\
&&\times\langle L_B M_{L_B} S_BM_{S_B}|J_BM_{J_B}\rangle\langle L_C M_{L_C} S_CM_{S_C}|J_CM_{J_C}\rangle\nonumber\\
&&\times\langle 1m1-m|00\rangle\langle \chi^{14}_{S_BM_{S_B}}\chi^{32}_{S_CM_{S_C}}|\chi^{12}_{S_AM_{S_A}}\chi^{34}_{1-m}\rangle\nonumber\\
&&\times[f_1I(\boldsymbol{P},m_1,m_2,m_3)\nonumber\\
&&+(-1)^{1+S_A+S_B+S_C}f_2I(-\boldsymbol{P},m_2,m_1,m_3)],
\label{helicity}
\end{eqnarray}
with $f_1=\langle \phi^{14}_B\phi^{32}_C|\phi^{12}_A\phi^{34}_0\rangle$ and $f_2=\langle \phi^{32}_B\phi^{14}_C|\phi^{12}_A\phi^{34}_0\rangle$, and
\begin{eqnarray}
I(\boldsymbol{P},m_1,m_2,m_3)=&&\int d^3\boldsymbol{p}\psi^*_{n_BL_BM_{L_B}}\left({\scriptstyle
\frac{m_3}{m_1+m_3}}\boldsymbol{P}_B+\boldsymbol{p}\right)\nonumber\\&&\times\psi^*_{n_CL_CM_{L_C}}\left({\scriptstyle
\frac{m_3}{m_2+m_3}}\boldsymbol{P}_B+\boldsymbol{p}\right)\nonumber\\&&\times\psi_{n_AL_AM_{L_A}}\left(\boldsymbol{P}_B+\boldsymbol{p}\right){\cal{Y}}^m_1(\boldsymbol{p}),
\label{overlap space}
\end{eqnarray}
where $\boldsymbol{P}={\boldsymbol{P}}_B=-{\boldsymbol{P}}_C$, $\boldsymbol{p}=\boldsymbol{p}_3$, $m_3$ is the mass of the created quark $q_3$.

The partial wave amplitude ${\cal{M}}^{LS}(\boldsymbol{P})$ can be given by~\cite{Jacob:1959at},
\begin{eqnarray}
{\cal{M}}^{LS}(\boldsymbol{P})&=&
\sum_{\renewcommand{\arraystretch}{.5}\begin{array}[t]{l}
\scriptstyle M_{J_B},M_{J_C},\\\scriptstyle M_S,M_L
\end{array}}\renewcommand{\arraystretch}{1}\!\!
\langle LM_LSM_S|J_AM_{J_A}\rangle \nonumber\\
&&\langle
J_BM_{J_B}J_CM_{J_C}|SM_S\rangle\nonumber\\
&&\times\int
d\Omega\,\mbox{}Y^\ast_{LM_L}{\cal{M}}^{M_{J_A}M_{J_B}M_{J_C}}
(\boldsymbol{P}). \label{pwave}
\end{eqnarray}

Various $^3P_0$ models exist in literature and typically differ in the choices of the pair-production vertex, the phase space conventions, and the meson wave functions employed. In this work, we restrict to the simplest vertex as introduced originally by Micu~\cite{Micu:1968mk} which assumes a spatially  constant pair-production strength $\gamma$, adopt the relativistic phase space, and employ two types of meson space wave functions: the simple  harmonic oscillator (SHO) wave functions and the relativized quark model (RQM) wave functions~\cite{Godfrey:1985xj}.

With the relativistic phase space, the decay width
$\Gamma(A\rightarrow BC)$ can be expressed in terms of the partial wave amplitude
\begin{eqnarray}
\Gamma(A\rightarrow BC)= \frac{\pi
|\boldsymbol{P}|}{4M^2_A}\sum_{LS}|{\cal{M}}^{LS}(\boldsymbol{P})|^2, \label{width1}
\end{eqnarray}
where $|\boldsymbol{P}|=\sqrt{[M^2_A-(M_B+M_C)^2][M^2_A-(M_B-M_C)^2]}{2M_A}$,
and $M_A$, $M_B$, and $M_C$ are the masses of the mesons $A$, $B$,
and $C$, respectively.

The parameters used in the $^3P_0$ model calculations involve the $q\bar{q}$ pair production strength $\gamma$,
the parameters associated with the meson wave functions, and the constituent quark masses. In the SHO wave functions case (case A), we follow the parameters used in Ref.~\cite{Close:2005se}, where the SHO wave function scale is $\beta=\beta_A=\beta_B=\beta_C=0.4$ GeV, the constituent quark masses are $m_u=m_d=330$~MeV, $m_s=550$~MeV, and $\gamma=8.77$ obtained by fitting to 32 well-established decay modes. In the RQM wave functions case (case B), we take $m_u=m_d=220$~MeV, and $m_s=419$~MeV as used in the relativized quark model of Godfrey and Isgur~\cite{Godfrey:1985xj}, and $\gamma=15.28$ by fitting to the same decay modes used in Ref.~\cite{Close:2005se} except for three decay modes without the specific branching ratios $K^{\ast\prime}\rightarrow \rho K$, $K^{\ast\prime}\rightarrow K^\ast\pi$, and $a_2\rightarrow \rho\pi$~\cite{PDG2016}. The meson flavor wave functions follow the conventions of Refs.~\cite{Godfrey:1985xj,Barnes:2002mu}. We assume that the $a_0(1450)$, $K^\ast_0(1430)$, $f_0(1370)$, and $f_0(1710)$ are the ground scalar mesons as in
Refs.~\cite{Barnes:2002mu,Ackleh:1996yt,Barnes:1996ff}. Masses of the final state mesons are taken from Ref.~\cite{PDG2016}.

\section{Results and discussions}
\label{sec:result}

\subsection{$\pi(2360)$}

\begin{figure}[htpb]
\includegraphics[scale=0.7]{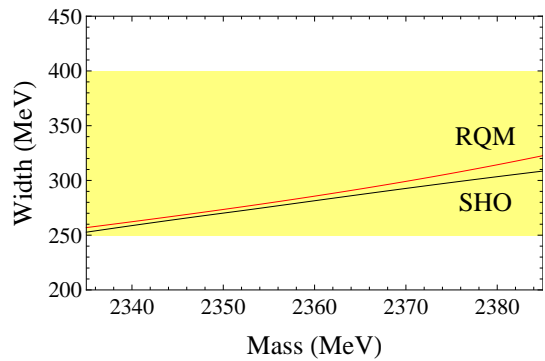}
\vspace{0.0cm}\caption{The dependence of the $\pi(5^1S_0)$ total width on the initial state mass in the $^3P_0$ model with two types of wave functions. The yellow band denotes the measured width for the $\pi(2360)$~\cite{Anisovich:2001pn,Anisovich:2001pp}.}\label{Fig:2360}
\end{figure}

The decay widths of the $\pi(2360)$ as the $\pi(5^1S_0)$ are listed in Table~\ref{tab:2360}. The $\pi(5^1S_0)$ total width is predicted to be about 281~MeV in case A or 285~MeV in case B, both in agreement with the $\pi(2360)$ width of $\Gamma=300^{+100}_{-50}$~MeV~\cite{Anisovich:2001pn,Anisovich:2001pp}. The dependence of the $\pi(5^1S_0)$ width on the initial mass is shown in Fig.~\ref{Fig:2360}. Within the $\pi(2360)$
mass errors ($2360\pm 25$~MeV), in both cases, the predicted width of the $\pi(5^1S_0)$ always overlaps with the $\pi(2360)$ width. Therefore, the measured width for the $\pi(2360)$ supports that the $\pi(2360)$ can be identified as the $\pi(5^1S_0)$. The flux-tube model calculations in Ref.~\cite{Wang:2017iai} also favor this assignment.

It is noted that for some decay modes such as $\pi\rho$, $\pi\rho(1700)$, $\pi(1300)\rho$, $\rho h_1(1170)$
$\omega b_1(1235)$, $\pi\rho_3(1690)$, and $K K^*_3(1780)$, the predictions in case A are similar with those in case B, while for other modes such as the $\pi f_0(1370)$, $\eta a_0(1450)$, $K K^\ast_0(1430)$, $\pi \rho(1450)$, $K K^*(1410)$, $K K^\ast(1680)$, $K(1460)K^\ast$, $\pi f_2(1270)$, $K K^\ast_2(1430)$, $\rho a_2(1320)$, and $\pi \rho_3(1990)$, there are some big variations between cases A and B. The similar behavior also exists in the flux-tube model (a variant of the $^3P_0$ model) calculations with different space wave functions~\cite{Blundell:1995ev,Kokoski:1985is}.

 As shown in Eqs. (\ref{helicity}) and (\ref{overlap space}), the partial width from the $^3P_0$ model depends on the overlap integrals of flavor, spin, and space wave functions of initial and final states. For a given decay mode, the overlap integrals of the flavor and spin wave functions of initial and final mesons are identical in both RQM and SHO cases, therefore, the partial width difference between the RQM and SHO cases results from the different choices of meson space wave functions. Generally speaking, the different space wave functions would lead to different decay widths. Especially, if the overlap is near to the nodes of space wave functions, the decay width would strongly depend on the details of wave functions, and the small wave function difference could generate a large discrepancy of the decay width. However, for some modes, the possibility that the different wave functions can give the similar decay widths also exists. To our knowledge, there is no some rules to judge whether the RQM and SHO wave functions can give the similar or different results before the numerical calculations.

The difference between the predictions in case A and those in case B provides a chance to distinguish among different meson space wave functions. At present, we are unable to conclude which type of wave function is more reasonable due to the lack of the branching ratios for the $\pi(2360)$. However, as suggested by Ref.~\cite{Lu:2016mbb}, we should keep in mind that it is essential to treat the wave functions accurately in the $^3P_0$ model calculations.

\begin{table}
\begin{center}
\caption{ \label{tab:2360} Decay widths of the $\pi(2360)$ as the $\pi(5^1S_0)$ with two types of wave functions (in MeV). The initial state mass is set to 2360~MeV. }
%\footnotesize
\begin{tabular}{c|ccc}
\hline\hline
 \multirow{2}*{\minitab[c]{Channel}}       & \multirow{2}*{\minitab[c]{Mode}}             & \multicolumn{2}{c}{$\Gamma_i$}\\
                                           &                              &SHO                &RQM \\\hline
  $0^-\rightarrow 0^-0^+$                   & $ \pi f_0(1370) $            & 1.31              & 44.22 \\
                                            & $\eta a_0(1450) $            & 0.17              & 11.07 \\
                                            & $K K^*_0(1430) $             & 0.08              & 7.03 \\
  \hline
  $0^-\rightarrow 0^-1^- $                  & $\pi\rho$                  & 1.67              & 1.67 \\
                                            & $\pi\rho(1450)$            & 0.002            & 25.98 \\
                                            & $\pi\rho(1700)$            & 2.35              & 2.83 \\
                                            & $\pi(1300)\rho$            & 23.54             & 29.81 \\
                                            & $K K^*$                     & 0.16              & 5.11 \\
                                            & $K K^*(1410)$               & 25.01             & 0.80\\
                                            & $K K^*(1680)$               & 2.10              &  0.0006\\
                                            & $K(1460) K^*$               & 0.77              & 0.01\\
  \hline
  $0^-\rightarrow 1^-1^+ $                  & $\rho a_1(1260)$            & 13.58             & 34.96\\
                                            & $\rho h_1(1170)$            & 10.64             & 17.45\\
                                            & $\omega b_1(1235) $          & 10.95             & 8.73\\

                                            & $K^* K_1(1270)$             & 8.49              & 0.08\\
                                            & $K^* K_1(1400)$             & 11.70             & 3.50\\
  \hline
  $0^-\rightarrow 1^-1^- $                  & $\rho \omega$               & 2.05              & 5.31\\
                                            & $\rho \omega(1420)$         & 37.15              & 5.29\\
                                            & $\omega\rho(1450)$          & 36.95             & 5.61\\
                                            & $K^* K^*$                   & 0.40              & 4.75\\
                                            & $K^* K^*(1410)$             & 20.64             & 0.69\\
  \hline
  $0^-\rightarrow 0^-2^+ $                  & $\pi f_2(1270) $             & 1.61              & 24.08\\
                                            & $\eta a_2(1320) $              & 4.07              & 2.99\\
                                            & $\eta^\prime a_2(1320) $       & 0.93              & 0.47\\

                                            & $K K^*_2(1430) $             & 12.67             & 0.0004\\
  \hline
  $0^-\rightarrow 1^-2^+ $                  & $\rho a_2(1320)$            & 27.57             & 3.17\\
                                            & $K^* K^*_2(1430)$           & 0.32              & 0.06\\
  \hline
  $0^-\rightarrow 0^-3^- $                  &$\pi\rho_3(1690)$           & 21.09             & 21.49\\
                                            &$\pi\rho_3(1990)$           & 3.42              & 18.31\\
                                            &$K K^*_3(1780)$              & 0.05              & 0.07\\
  \hline
  $0^-\rightarrow 0^-4^+ $                  &$\pi f_4(2050) $           & 0.01            & 0.12\\\hline
 \multicolumn{2}{c|}{Total width}        &  281.46    & 285.65\\
  \hline
  \multicolumn{2}{c|}{Experiment}  & \multicolumn{2}{c}{$300^{+100}_{-50}$~\cite{Anisovich:2001pn,Anisovich:2001pp}}\\
\hline\hline
\end{tabular}
\end{center}
\end{table}

\subsection{$\eta(2320)$ and $X(2500)$ }
\label{sec:eta2320}

The $\eta(2320)$ was observed in $\bar{p}p\rightarrow \eta\eta\eta$ process, and its mass and width are $2320\pm 15$ MeV and $230\pm 35$ MeV~\cite{Anisovich:2000ix}. The predicted $\eta(5^1S_0)$ mass in the relativistic quark model is about $2385$ MeV~\cite{Ebert:2009ub}, close to the $\eta(2320)$ mass. In the presence of the $X(2500)$ as the isoscalar member of the $5^1S_0$ meson nonet~\cite{Pan:2016bac}, we shall discuss the possibility of the $\eta(2320)$ as the isoscalar partner of the $X(2500)$.

In a meson nonet, the two physical isoscalar states can mix. The mixing of the two isoscalar states can be parametrized as
\begin{eqnarray}
&&\eta(5^1S_0)=\cos\phi~ n\bar{n}-\sin\phi~ s\bar{s},\\
&&X(2500)=\sin\phi~ n\bar{n}+\cos\phi~ s\bar{s},
\end{eqnarray}
where $n\bar{n}=(u\bar{u}+d\bar{d})/\sqrt{2}$ and $s\bar{s}$ are
the pure $5\,^1S_0$ nonstrange and strange states, respectively, and the $\phi$ is the mixing angle. Accordingly, the partial widths for the $\eta(5^1S_0)$ and
$X(2500)$ can be expressed as
\begin{eqnarray}
\Gamma(\eta(5^1S_0)\rightarrow
BC)&=&\frac{\pi~P}{4M^2_{\eta(5^1S_0)}}\sum_{LS}|\cos\phi
{\cal{M}}^{LS}_{n\bar{n}\rightarrow BC}\nonumber \\ && -\sin\phi
{\cal{M}}^{LS}_{s\bar{s}\rightarrow BC}|^2,
\label{w1}\\
 \Gamma(X(2500)\rightarrow
BC)&=&\frac{\pi~P}{4M^2_{X(2500)}}\sum_{LS}|\sin\phi
{\cal{M}}^{LS}_{n\bar{n}\rightarrow BC} \nonumber \\ &&
+\cos\phi
{\cal{M}}^{LS}_{s\bar{s}\rightarrow BC}|^2. \label{w2}
\end{eqnarray}

\begin{table*}[htb]
\begin{center}
\caption{ \label{tab:mixing1} Decay widths of the $\eta(2320)$ and $X(2500)$ as the $5^1S_0$  isoscalar states with the SHO wave functions (in MeV). $c\equiv\cos\phi,s\equiv\sin\phi.$ The masses of the $\eta(2320)$ and $X(2500)$ are set to 2320 and 2470 MeV, respectively~\cite{Ablikim:2016hlu,Anisovich:2000ix}. A dash indicates that a decay mode is forbidden.}
%\footnotesize
\begin{tabular}{c|ccc}
\hline\hline
  \multirow{2}*{\minitab[c]{Channel}}  & \multirow{2}*{\minitab[c]{Mode}}     &$\eta(2320)$      & $X(2500)$ \\
                                     &                        &$\Gamma_i$        &$\Gamma_i$ \\   \hline
  $0^-\rightarrow 0^-0^+$            & $\pi a_0(1450) $       & $3.69  c^2 $     & $3.84 s^2 $   \\
                                     & $\eta f_0(1370) $      & $0.40 c^2 $      & $0.79 s^2   $   \\
                                     & $\eta^\prime f_0(1370)$ & $-$             & $0.22 s^2   $   \\
                                     & $\eta f_0(1710) $      & $2.48 s^2 $      & $0.20 c^2   $ \\
      &$K K^*_0(1430) $        & $0.0006  c^2 -0.07cs+2.25s^2  $          & $1.17 c^2 +1.98cs+0.84s^2  $  \\ \hline

  $0^-\rightarrow 0^-1^- $     & $K K^*$  & $0.06  c^2 +0.35cs+0.51s^2 $ & $0.06  c^2 +0.39cs+0.65s^2  $   \\
                               & $K K^*(1680)$    &  $1.51 c^2 +5.40cs+4.82s^2 $  & $2.29  c^2 -5.11cs+2.85s^2  $  \\
                               & $K K^*(1410)$  &  $27.47  c^2 -3.84cs+0.13s^2 $ & $13.83 c^2 +29.02cs+15.23s^2  $  \\
                               & $K(1460) K^*$  &  $-$    & $24.94  c^2 -54.68cs+25.40s^2  $   \\
                               & $K K^*(1830)$  &  $-$    & $72.21 c^2 -63.56cs+135.49s^2  $  \\  \hline
  $0^-\rightarrow 0^-2^+ $     & $\pi a_2(1320) $    & $10.99 c^2 $   & $0.85 s^2   $\\
                               & $\eta f_2(1270) $   & $4.05  c^2   $  & $1.60s^2   $\\
                               & $\eta^\prime f_2(1270) $    & $0.96 c^2 $   & $4.86 s^2 $\\
                               & $\eta f^\prime_2(1525) $    & $6.53 s^2 $   & $9.61c^2  $\\
                               & $K K^*_2(1430) $    & $10.90 c^2 -18.89cs+8.19s^2 $  & $0.43 c^2 +5.34cs+16.51s^2$ \\ \hline
  $0^-\rightarrow 0^-3^-$ &$K K^*_3(1780)$    & $0.007  c^2 +0.05cs+0.10s^2 $ & $9.25  c^2 -5.95cs+0.96s^2  $ \\\hline
  $0^-\rightarrow 0^-4^+ $  & $\pi a_4(2040) $  & $0.02  c^2$             & $0.83 s^2   $ \\ \hline
  $0^-\rightarrow 1^-1^- $  & $\rho \rho$       & $2.63 c^2  $           & $3.78 s^2   $ \\
                           & $\rho \rho(1450)$  & $84.95 c^2 $        & $100.69 s^2   $\\
                           & $\omega \omega$    & $0.83 c^2 $         & $1.26 s^2   $\\
                           & $\omega \omega(1420)$   & $24.43 c^2 $    & $35.69s^2   $ \\
                           & $\phi\phi$         & $1.16 s^2 $          & $0.01c^2   $ \\
                           & $K^* K^*$     & $0.63 c^2 +1.97cs+0.54s^2 $      & $1.37  c^2 -0.35cs+0.02s^2  $\\
                           & $K^* K^*(1410)$ & $3.30c^2 -11.29cs+9.65s^2 $  & $42.44c^2 +104.60cs+64.44s^2  $\\\hline
  $0^-\rightarrow 1^-1^+ $ & $\rho b_1(1235)$ & $31.05 c^2 $        & $28.29 s^2   $\\
                           &$\omega h_1(1170)$ & $10.87 c^2 $        & $7.57 s^2   $\\
                           & $K^* K_1(1270)$   & $5.34 c^2 -6.82cs+2.22s^2$      & $8.76  c^2 +11.89cs+18.55s^2  $\\
     & $K^* K_1(1400)$   & $10.88  c^2 +9.29cs+2.51s^2 $   & $18.13  c^2 +0.85cs+4.14s^2  $\\  \hline
  $0^-\rightarrow 1^-2^+ $   &$K^* K^*_2(1430)$    & $0.0004c^2 +0.002cs+0.003s^2$  & $18.59 c^2 -20.81cs+5.83s^2  $\\
  \hline
 \multicolumn{2}{c|}{Total width}        &   $234.93  c^2 -10.07cs+42.09s^2  $     &  $227.34  c^2 +3.70cs+481.18s^2  $ \\
  \hline
  \multicolumn{2}{c|}{Experiment} & $230\pm 35$ ~\cite{Anisovich:2000ix}    & $230^{+64+56}_{-35-33}$ ~\cite{Ablikim:2016hlu}                                  \\
\hline\hline
\end{tabular}
\end{center}
\end{table*}

\begin{table*}[htb]
\begin{center}
\caption{ \label{tab:mixing2} Decay widths of the $\eta(2320)$ and $X(2500)$ as the $5^1S_0$  isoscalar states with the RQM wave functions (in MeV). $c\equiv\cos\phi,s\equiv\sin\phi$. The masses of the $\eta(2320)$ and $X(2500)$ are set to 2320 and 2470 MeV, respectively~\cite{Ablikim:2016hlu,Anisovich:2000ix}. A dash indicates that a decay  mode is forbidden.}
%\footnotesize
\begin{tabular}{c|ccc}
\hline\hline
   \multirow{2}*{\minitab[c]{Channel}}   & \multirow{2}*{\minitab[c]{Mode}}   &$\eta(2320)$  & $X(2500)$ \\
                                           &     &$\Gamma_i$                &$\Gamma_i$ \\  \hline
  $0^-\rightarrow 0^-0^+$     & $\pi a_0(1450) $    & $93.36 c^2 $    & $141.51 s^2   $   \\
                             & $\eta f_0(1370) $    & $13.30 c^2 $    & $21.64 s^2   $   \\
                             & $\eta^\prime f_0(1370) $   & $-$       & $8.02 s^2   $   \\
                             & $\eta f_0(1710) $   & $1.46s^2 $    & $3.42 c^2   $ \\
   & $K K^*_0(1430) $    & $6.33c^2 -34.96cs+48.24s^2  $          & $84.74 c^2 +55.93cs+9.23s^2  $  \\  \hline
  $0^-\rightarrow 0^-1^- $       & $K K^*$       &  $4.81 c^2 -4.59cs+1.10s^2$     & $0.97  c^2 +4.78cs+5.85s^2  $ \\
  & $K K^*(1680)$               &  $0.003 c^2 +0.01cs+0.02s^2  $           &  $4.81  c^2 -0.58cs+0.02s^2  $  \\
  & $K K^*(1410)$               &  $0.49  c^2 +6.77cs+23.54s^2  $          & $10.53 c^2 -9.13cs+1.98s^2  $  \\
  & $K(1460) K^*$               &  $-$                                    & $0.08  c^2 +0.02cs+0.001s^2  $   \\
  & $K K^*(1830)$               &  $-$                              & $38.40 c^2 -32.26cs+6.78s^2$  \\  \hline
  $0^-\rightarrow 0^-2^+ $      & $\pi a_2(1320) $        & $50.99 c^2   $                     & $88.50 s^2   $\\
                                & $\eta f_2(1270)$        & $2.88  c^2   $                     & $10.95s^2   $\\
                   & $\eta^\prime f_2(1270) $  & $0.47c^2 $          & $0.04 s^2   $\\
                   & $\eta f^\prime_2(1525) $    & $0.01 s^2 $                & $0.12c^2   $\\
                   & $K K^*_2(1430) $        & $0.01  c^2 +0.56cs+6.22s^2 $        & $24.99 c^2 +3.60cs+0.13s^2  $\\
    \hline
  $0^-\rightarrow 0^-3^- $               & $K K^*_3(1780)$             & $0.004 c^2 +0.10cs+0.55s^2  $           & $15.06 c^2 -4.27cs+0.30s^2  $ \\
  \hline
  $0^-\rightarrow 0^-4^+ $               & $\pi a_4(2040) $             & $0.22  c^2   $                           & $6.69 s^2   $ \\
  \hline
  $0^-\rightarrow 1^-1^- $               & $\rho \rho$                & $11.20 c^2   $                            & $0.67 s^2   $ \\
                                         & $\rho \rho(1450)$          & $29.40 c^2   $                           & $32.06 s^2   $\\
                                         & $\omega \omega$            & $4.07 c^2   $                            & $0.33 s^2   $\\
                                         & $\omega \omega(1420)$      & $10.48 c^2   $                           & $7.14s^2   $ \\
                                         & $\phi\phi$                 & $1.10 s^2   $                            & $1.71c^2   $ \\
                                         & $K^* K^*$                  & $4.25  c^2 -1.47cs+0.13s^2  $            & $3.23  c^2 -8.78cs+5.96s^2  $\\
                                         & $K^* K^*(1410)$            & $0.16  c^2 -2.28cs+7.86s^2  $           & $8.59  c^2 -2.35cs+0.16s^2  $\\
  \hline
  $0^-\rightarrow 1^-1^+ $               & $\rho b_1(1235)$            & $17.12 c^2   $                           & $96.26 s^2   $\\
                                         & $\omega h_1(1170)$          & $10.12 c^2   $                           & $45.91 s^2   $\\
  & $K^* K_1(1270)$             & $0.09 c^2 -2.07cs+12.55s^2  $         & $53.22 c^2 +0.48cs+0.001s^2  $\\
  & $K^* K_1(1400)$             & $2.54  c^2 -2.24cs+0.54s^2  $         & $0.56  c^2 +3.12cs+4.41s^2  $\\
  \hline
  $0^-\rightarrow 1^-2^+ $               &$K^* K^*_2(1430)$            & $0.00007  c^2 +0.001cs+0.007s^2  $        & $7.26 c^2 -4.08cs+0.47s^2  $\\
  \hline
 \multicolumn{2}{c|}{Total width}        &$262.31 c^2 -40.16cs+103.32s^2  $     &  $257.71  c^2 +6.48cs+495.09s^2  $ \\
  \hline
  \multicolumn{2}{c|}{Experiment} & $230\pm 35$~\cite{Anisovich:2000ix}    & $230^{+64+56}_{-35-33}$~\cite{Ablikim:2016hlu}                                  \\
\hline\hline
\end{tabular}
\end{center}
\end{table*}

\begin{figure}[hbt]
\includegraphics[scale=0.7]{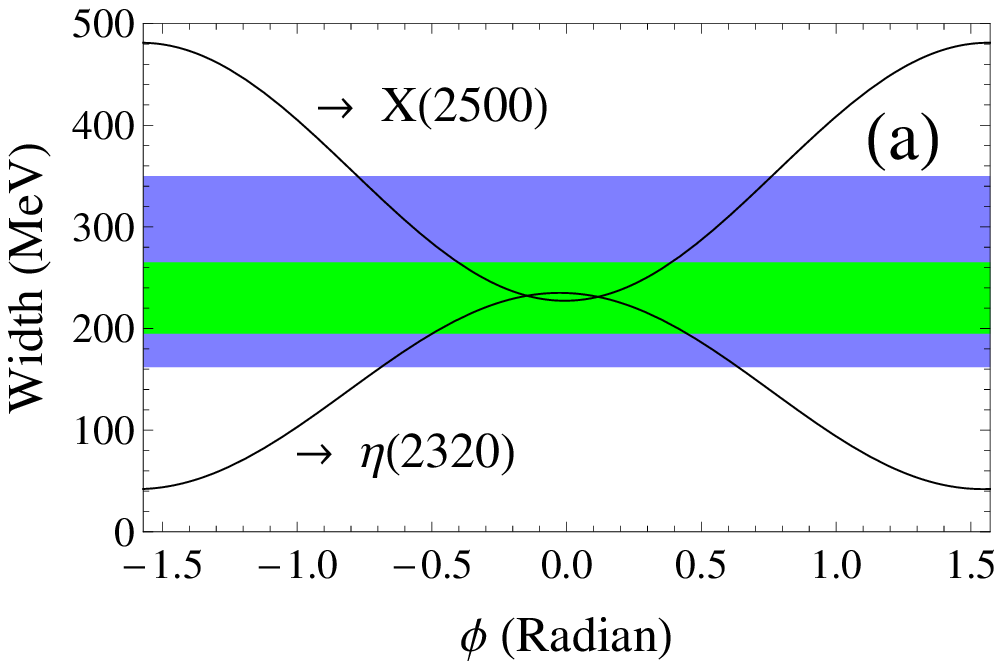}
\includegraphics[scale=0.7]{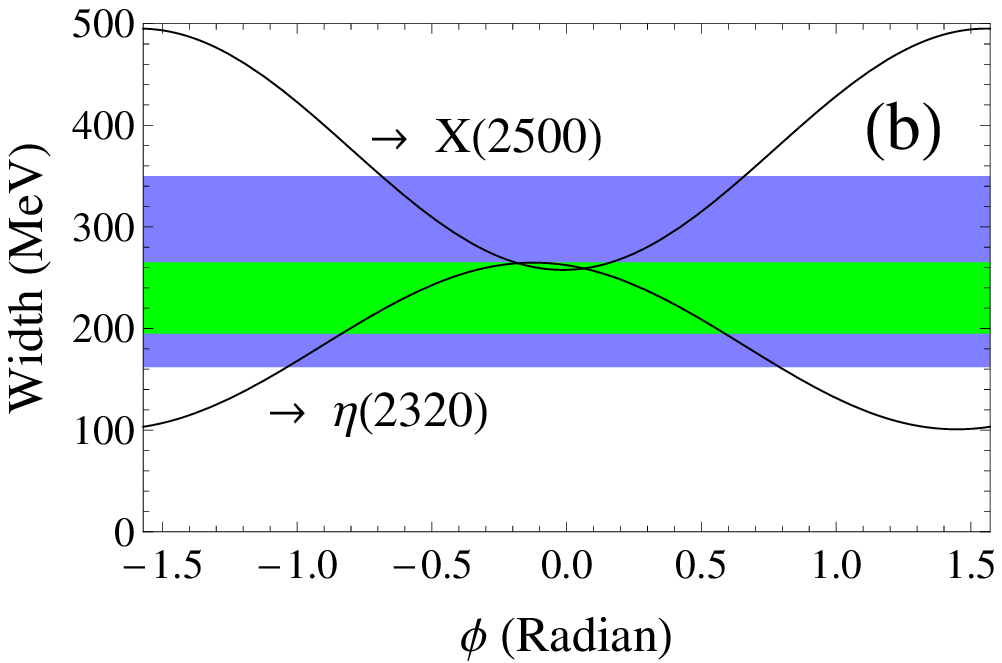}
\vspace{0.0cm}\caption{The dependence of the $\eta(2320)$ and $X(2500)$ total widths on the $\phi$ in the $^3P_0$ model with two types of wave functions: (a) with the SHO wave functions (b) with the RQM wave functions. The blue and green bands denote the measured widths for the $X(2500)$ and $\eta(2320)$, respectively~\cite{Ablikim:2016hlu,Anisovich:2000ix}. }\label{Fig:eta2320}
\end{figure}

Under the mixing of $\eta(2320)$ and $X(2500)$, their decays in the case A are listed in  Table~\ref{tab:mixing1} and those in the case B are listed in Table~\ref{tab:mixing2}. The dependence of the $\eta(2320)$ and $X(2500)$ total widths on the mixing angle $\phi$ is displayed in Fig.~\ref{Fig:eta2320}. In order to simultaneously reproduce the measured widths for the $\eta(2320)$ and $X(2500)$, the mixing angle $\phi$ is required to satisfy  $-0.5\leq \phi \leq 0.45$ radians in case A or $-0.69\leq \phi \leq 0.59$ radians in case B. Below, we shall estimate the value of $\phi$ to check whether it satisfies these constraints based on the mass-squared describing the mixing of two isoscalar mesons.

In the $n\bar{n}$ and $s\bar{s}$ bases, the mass-squared matrix describing the $\eta(2320)$ and $X(2500)$ mixing can
be expressed as~\cite{Li:2008mza,Li:2008et,Li:2001qg}

\begin{eqnarray}
M^2=\left(\begin{array}{cc}
M^2_{n\bar{n}}+2A_m&\sqrt{2}A_mX\\
\sqrt{2}A_mX&M_{s\bar{s}}^2+A_mX^2
\end{array}\right),
\label{matrix}
\end{eqnarray}
where $M_{n\bar{n}}$ and $M_{s\bar{s}}$ are the masses of the pure
$5^1S_0$ $n\bar{n}$ and $s\bar{s}$, respectively, $A_m$ denotes the
total annihilation strength of the $q\bar{q}$ pair for the light
flavors $u$ and $d$, $X$ describes the SU(3)-breaking ratio of
the nonstrange and strange quark masses via the constituent quark
mass ratio $m_u/m_s$. Since the $n\bar{n}$ is the orthogonal partner of the $\pi(5^1S_0)$, one can expect that $n\bar{n}$ degenerates with $\pi(5\,^1S_0)$ in effective quark
masses. Here we take $M_{n\bar{n}}=M_{\pi(5^1S_0)}=M_{\pi(2360)}$. The $M_{s\bar{s}}$ can be obtained from the Gell-Mann-Okubo mass formula $M^2_{s\bar{s}}=2M^2_{ K(5^1S_0)}-M^2_{n\bar{n}}$.

The masses of the two physical states
$\eta(2320)$ and $X(2500)$ can be related to the matrix $M^2$
by the unitary matrix
\begin{eqnarray}
 U=\left(
 \begin{array}{cc}
\cos\phi&-\sin\phi\\
\sin\phi&\cos\phi
\end{array}\right),
\end{eqnarray}
which satisfies
\begin{eqnarray}
U M^2 U^\dagger=\left(\begin{array}{cc}
M^2_{\eta(2320)}&0\\
0&M^2_{X(2500)}\end{array}\right). \label{diag}
\end{eqnarray}

From Eqs.~(\ref{matrix}) and (\ref{diag}), one can have
\begin{eqnarray}
&& 8X^2(M^2_{K(5^1S_0)}-M^2_{\pi(2360)})^2 \nonumber \\
&=&\left[4M^2_{K(5^1S_0)}-(2-X^2)M^2_{\pi(2360)}-(2+X^2)M^2_{\eta(2320)}\right]\nonumber\\
&\times&\left[(2-X^2)M^2_{\pi(2360)}+(2+X^2)M^2_{X(2500)}-4M^2_{K(5^1S_0)}\right], \nonumber \\
\label{msch}
\end{eqnarray}
and
\begin{eqnarray}
A_m&=&(M^2_{X(2500)}-2M^2_{K(5^1S_0)}+M^2_{\pi(2360)})(M^2_{\eta(2320)}-2M^2_{K(5^1S_0)} \nonumber \\ &&  +M^2_{\pi(2360)})/\left[2(M^2_{\pi(2360)}-M^2_{K(5^1S_0)})X^2\right].
\label{am}
\end{eqnarray}
Equation (\ref{msch}) is the generalized Schwinger's nonet mass formula~\cite{Li:2001qg}. If the SU(3)-breaking effect is not considered, i.e., $X=1$, Eq.~(\ref{msch}) can be reduced to original Schwinger's nonet mass formula~\cite{Schwinger:1964zza}. With the masses of the $\pi(2360)$, $\eta(2320)$, and $X(2500)$, from Eqs.~(\ref{msch}) and (\ref{am}), we have
\begin{eqnarray}
M_{K(5\,^1S_0)}=2.418~\mbox{GeV}, A_m=-0.085~ {\mbox{GeV}}^2
\label{eq: Kmass1}
\end{eqnarray}
for $X=m_u/m_s=330/550$ as used in case A, and
\begin{eqnarray}
M_{K(5\,^1S_0)}=2.418~ \mbox{GeV}, A_m=-0.111~ {\mbox{GeV}}^2
\label{eq:Kmass2}
\end{eqnarray}
for $X=m_u/m_s=220/419$ as used in case B.

Then the unitary matrix $U$ can be given by
\begin{eqnarray}
U=\left(\begin{array}{cc}
\cos\phi&-\sin\phi\\
\sin\phi&\cos\phi\end{array}\right)=\left(\begin{array}{cc}
 +0.995&+0.102\\
-0.102&+0.995
\end{array}\right)
\label{mixangle1}
\end{eqnarray}
for $X=330/550$, and
\begin{eqnarray}
U=\left(\begin{array}{cc}
\cos\phi&-\sin\phi\\
\sin\phi&\cos\phi\end{array}\right)=\left(\begin{array}{cc}
 +0.994&+0.109\\
-0.109&+0.994
\end{array}\right)
\label{mixangle2}
\end{eqnarray}
for $X=220/419$.

Equations (\ref{mixangle1}) and (\ref{mixangle2}) consistently give $\phi=-0.1$ radians, which makes both the $\eta(2320)$ and $X(2500)$ widths in agreement with experimental data. Also, both Eqs.~(\ref{mixangle1}) and (\ref{mixangle2}) indicate that the $\eta(2320)$ is mainly the $n\bar{n}$, consistent with the $\pi(2360)$ nearly degenerating with the $\eta(2320)$, while the $X(2500)$ is mainly the $s\bar{s}$, consistent with our previous analysis~\cite{Pan:2016bac}. Therefore, the $\eta(2320)$ and $X(2500)$, together with the $\pi(2360)$, appear to be the convincing $5\,^1S_0$ states.

In above discussions, we focus on the possibility of the pseudoscalar states $\pi(2360)$, $\eta(2320)$, and $X(2500)$ as the $5^1S_0$ mesons.  Apart from the states
listed in Table~\ref{tab:pseu}, the $X(2120)$ and $X(2370)$ also probably are the $J^{PC}=0^{-+}$ resonances. The $X(2120)$ and $X(2370)$ were observed by the BESIII collaboration in the $\pi^+\pi^-\eta^\prime$ invariant mass spectrum and their spin parities are not determined~\cite{Ablikim:2010au}. Based on the observed decay mode $\pi^+\pi^-\eta^\prime$, the possible $J^{PC}$ for the $X(2120)$ and $X(2370)$ are $0^{-+}$, $1^{++}$, $\cdots$. The natures of the $X(2120)$ and $X(2370)$ are not clear~\cite{Wang:2017iai,Yu:2011ta,Liu:2010tr,Chen:2011kp,Wang:2010vz}. Since the $X(2370)$ mass is also close to the quark model expectation for the $\eta(5^1S_0)$ mass~\cite{Ebert:2009ub}, we shall discuss the possibility of the $X(2370)$ as the isoscalar partner of the $X(2500)$.

\begin{figure}[htpb]
\includegraphics[scale=0.7]{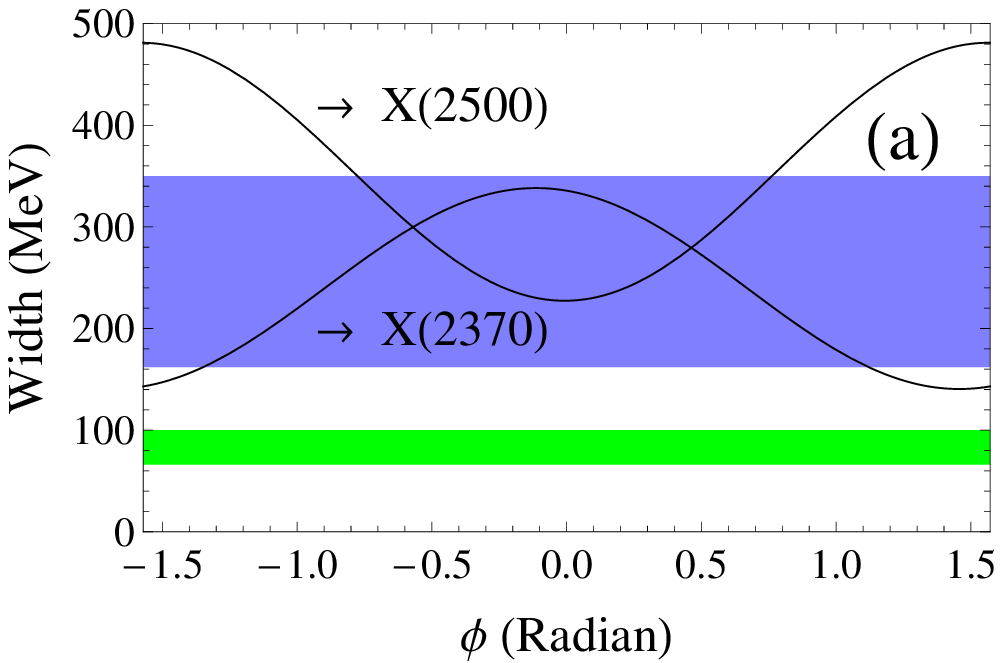}
\includegraphics[scale=0.7]{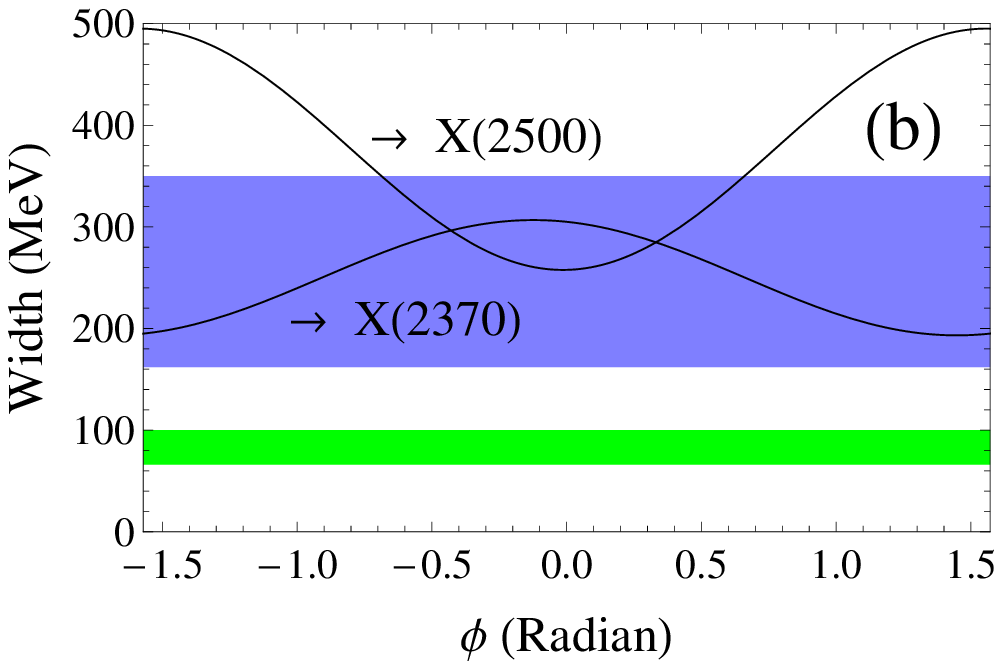}
\vspace{0.0cm}\caption{The total widths of the $\eta(2320)$ and $X(2500)$  dependence on the $\phi$ in the $^3P_0$ model with two types of wave functions: (a) with the SHO wave functions (b) with the RQM wave functions. The blue and green band denote the measured widths for the $X(2500)$ and $X(2370)$, respectively~\cite{Ablikim:2016hlu,Ablikim:2010au}.}\label{Fig:eta2370}
\end{figure}

With the $X(2370)$-$X(2500)$ mixing, the decay widths for the $X(2370)$ are listed in Table~\ref{tab:2370}. The dependence of the $X(2370)$ and $X(2500)$ total widths on the mixing angle is plotted in Fig.~\ref{Fig:eta2370}. Obviously, the $X(2370)$ width can not be reproduced in the whole region of the mixing angle. Therefore, our calculations do not support the $5^1S_0$ assignment for the $X(2370)$. Other calculations from the $^3P_0$ model suggest that the $X(2370)$ is unlikely to be the $4^1S_0$ $q\bar{q}$ state~\cite{Liu:2010tr,Chen:2011kp}. If the $X(2370)$ turns out to have $J^{PC}=0^{-+}$ in future, in order to explain its properties, more complicate scheme such as the $q\bar{q}$-glueball mixing  may be necessary, since the $X(2370)$ mass is close to the pseudoscalar glueball mass of about $2.3-2.6$ GeV predicted by the lattice QCD~\cite{Morningstar:1999rf,Hart:2001fp,Chen:2005mg}.

\begin{table*}
\begin{center}
\caption{ \label{tab:2370}Decay widths of the $X(2370)$ as the $5^1S_0$  state in the $^3P_0$ model with two types of wave functions (in MeV). The initial state mass is set to $2376.3$ MeV~\cite{Ablikim:2010au}.}
%\footnotesize
\begin{tabular}{c|ccc}
\hline\hline
\multirow{2}*{\minitab[c]{Channel}}       & \multirow{2}*{\minitab[c]{Mode}}             & \multicolumn{2}{c}{$\Gamma_i$}\\
                                          &                              &SHO                &RQM \\\hline
$0^-\rightarrow 0^-0^+$                   & $\pi a_0(1450) $            & $4.16  c^2   $                           & $110.41 c^2   $   \\
                                            & $\eta f_0(1370) $            & $0.62 c^2   $                            & $16.15 c^2   $   \\
                                            & $\eta^\prime f_0(1370) $     & $1.52 c^2   $                       &$4.73 c^2   $   \\
                                            & $\eta f_0(1710) $            & $1.63s^2   $                            & $2.22 s^2   $ \\
                                            & $K K^*_0(1430) $             & $0.15  c^2 -1.14cs+2.19s^2  $          & $7.33 c^2 -42.31cs+61.07s^2  $  \\
  \hline
  $0^-\rightarrow 0^-1^- $                  & $K K^*$                     &  $0.21 c^2 +0.36cs+0.16s^2  $           & $5.22  c^2 -5.22cs+1.30s^2  $   \\
                                            & $K K^*(1680)$               &  $2.29 c^2 +6.74cs+4.94s^2  $            &  $0.00006  c^2 -0.009cs+0.36s^2  $  \\
                                            & $K K^*(1410)$               &  $23.78  c^2 -17.56cs+3.24s^2  $          & $0.95 c^2 +9.21cs+22.38s^2  $  \\
                                           & $K(1460) K^*$               &   $3.67 c^2 +12.09cs+9.95s^2  $            & $0.002  c^2 +0.65cs+3.11s^2  $   \\
                                             & $K K^*(1830)$               &    $16.65 c^2 +48.04cs+34.66s^2  $      & $2.59 c^2 +21.80cs+45.82s^2  $  \\
  \hline
  $0^-\rightarrow 0^-2^+ $                   & $\pi a_2(1320) $            & $6.35 c^2   $                           & $67.47 c^2   $\\
                                            & $\eta f_2(1270) $            & $3.26  c^2   $                           & $5.52c^2   $\\
                                            & $\eta^\prime f_2(1270) $     & $2.47 c^2   $                            & $0.59 c^2   $\\
                                             & $\eta f^\prime_2(1525) $    & $8.40 s^2   $                            & $0.0002s^2   $\\
                                            & $K K^*_2(1430) $             & $13.36  c^2 -15.98cs+4.78s^2  $          & $0.0007 c^2 -0.19cs+13.13s^2  $ \\
  \hline
  $0^-\rightarrow 0^-3^- $                  & $K K^*_3(1780)$             & $0.10  c^2 +0.71cs+1.30s^2  $           & $0.04  c^2 +0.81cs+3.97s^2  $ \\
  \hline
  $0^-\rightarrow 0^-4^+ $                  & $\pi a_4(2040) $             & $0.11  c^2   $                           & $1.02 c^2   $ \\
  \hline
  $0^-\rightarrow 1^-1^- $                   & $\rho \rho$                & $3.29 c^2   $                            & $6.14 c^2   $ \\
                                             & $\rho \rho(1450)$          & $119.77 c^2   $                           & $6.80 c^2   $\\
                                             & $\omega \omega$            & $1.07 c^2   $                            & $2.33 c^2   $\\
                                             & $\omega \omega(1420)$      & $38.85 c^2   $                           & $3.87c^2   $ \\
                                             & $\phi\phi$                 & $0.43 s^2   $                            & $1.41s^2   $ \\
                                             & $K^* K^*$                  & $0.32  c^2 +1.49cs+1.74s^2  $            & $4.94  c^2 +1.86cs+0.18s^2  $\\
                                             & $K^* K^*(1410)$            & $28.51  c^2 -78.14cs+53.54s^2  $           & $0.79  c^2 -6.84cs+14.88s^2  $\\
  \hline
  $0^-\rightarrow 1^-1^+ $                  & $\rho b_1(1235)$            & $33.12 c^2   $                           & $34.57 c^2   $\\
                                            & $\omega h_1(1170)$          & $10.51 c^2   $                           & $19.70 c^2   $\\
                                            & $K^* K_1(1270)$             & $9.91 c^2 -10.15cs+5.03s^2  $             & $0.07  c^2 -2.30cs+18.79s^2  $\\
                                            & $K^* K_1(1400)$             & $10.91  c^2 +5.54cs+7.06s^2  $           & $3.71  c^2 -6.82cs+0.01s^2  $\\
  \hline
  $0^-\rightarrow 1^-2^+ $                  &$K^* K^*_2(1430)$            & $0.69  c^2 +3.30cs+3.98s^2  $        & $0.11 c^2 +1.69cs+6.39s^2  $\\
  \hline
 \multicolumn{2}{c|}{Total width}        &  $335.65  c^2 -44.70cs+143.03s^2  $    & $305.01 c^2 -27.66cs+195.00s^2  $   \\
  \hline
  \multicolumn{2}{c|}{Experiment}  & \multicolumn{2}{c}{$83\pm 17$ ~\cite{Ablikim:2010au}}\\
\hline\hline
\end{tabular}
\end{center}
\end{table*}

\subsection{$K(5^1S_0)$}
\label{subsec:dvector}

\begin{table}
\begin{center}
\caption{ \label{tab:2420}Decay widths of the $K(2418)$ as the $5^1S_0$ state in the $^3P_0$ model with two types of wave functions (in MeV). The initial state mass is set to 2418 MeV.}
%\footnotesize
\begin{tabular}{c|ccc}
\hline\hline
 \multirow{2}*{\minitab[c]{Channel}}       & \multirow{2}*{\minitab[c]{Mode}}             & \multicolumn{2}{c}{$\Gamma_i$}\\
                                           &                              &SHO                &RQM \\\hline
\hline
  $0^-\rightarrow 0^-0^+$                  & $\pi K^*_0(1430) $            & 0.36              & 32.62   \\
                                           & $K a_0(1450) $                & 0.96              & 19.08  \\
                                           & $\eta K^*_0(1430) $            & 0.03              & 0.36   \\
                                           & $\eta^\prime K^*_0(1430) $     & 2.12              & 4.71   \\
                                           & $K f_0(1370)$                 & 0.31              & 8.48  \\
                                            & $K f_0(1710)$                 & 0.33              & 1.64  \\
  \hline
  $0^-\rightarrow 0^-1^- $                 & $\pi K^* $                     & 0.07                &  0.05   \\
                                           & $K \rho $                     & 0.14              & 0.10 \\
                                           & $\pi K^*(1680) $               & 0.05              & 2.67 \\
                                           & $K \rho(1700)$                & 1.84              & 0.35 \\
                                           & $\pi K^*(1410) $               & 5.75              & 0.10 \\
                                           & $K \rho(1450)$                & 0.06              & 8.50 \\
                                           & $\pi(1300)K^*$               & 8.11              & 7.11 \\
                                           & $K(1460) \rho$                & 9.69              & 2.14\\
                                           & $\eta K^* $                    & 0.42              & 0.001\\
                                           & $\eta^\prime K^* $              & 0.02              & 0.10\\
                                           & $\eta K^*(1410) $              & 0.03              & 9.51\\
                                           & $\eta^\prime K^*(1410) $        & 0.21              & 0.04\\
                                           & $\eta K^*(1680) $              & 2.00              & 0.03\\
                                           & $K \phi$                & 0.36              & 0.93\\
                                           & $K \phi(1680)$                & 10.41              & 0.04\\
                                           & $\eta(1475)K^*$              & 5.17             &  0.03\\
                                           & $K \omega$                    & 0.05                & 0.03\\
                                           & $K \omega(1420)$              & 0.03              & 2.81\\
                                           & $\eta(1295)K^*$              & 2.54              & 2.61\\
                                           & $K \omega(1650)$              & 0.49              & 0.32\\
                                           & $K(1460) \omega$              & 3.39              & 0.58\\
  \hline
  $0^-\rightarrow 0^-2^+ $
                                           & $\pi K^*_2(1430) $             & 0.19              & 13.75\\
                                           & $K a_2(1320)$                 & 1.04              & 7.24\\
                                           & $\eta K^*_2(1430) $            & 0.13              &  0.04\\
                                           & $\eta^\prime K^*_2(1430) $      & 0.27              & 0.23\\
                                           & $K f_2^\prime(1525)$           & 5.74              & 0.05\\
                                           & $K f_2(1270)$                 &  0.15                & 2.86\\
  \hline
  $0^-\rightarrow 0^-3^- $                 & $\pi K^*_3(1780) $             & 9.77             & 6.14\\
                                           & $K \rho_3(1690)$              & 3.18              & 3.66\\
                                           & $\eta K^*_3(1780) $            & 0.28              & 0.91\\
                                           & $K \phi_3(1850)$              & 0.03              & 0.007\\
                                           & $K \omega_3(1670)$            & 1.36              & 1.42\\
  \hline
  $0^-\rightarrow 0^-4^+ $                 & $\pi K^*_4(2045) $             & 0.07              & 0.79\\
  \hline
  $0^-\rightarrow 1^-1^- $                 & $K^* \rho$                    & 1.47              &0.16 \\
                                           & $K^*(1410) \rho$              & 12.60              & 13.47\\
                                           & $K^* \rho(1450)$              & 19.21             & 6.47\\
                                           & $K^* \phi$              & 0.03              & 1.93\\
                                           & $K^* \omega$                  & 0.49              & 0.04\\
                                           & $K^*(1410) \omega$            & 4.55              & 3.91\\
                                           & $K^* \omega(1420)$            & 5.94              & 2.21\\
  \hline
  $0^-\rightarrow 1^-1^+ $                 &$K^* b_1(1235)$                & 6.66              & 4.15\\
                                           &$K^* a_1(1260)$                & 4.35              & 8.46\\
                                           &$\rho K_1(1270)$               & 3.47              & 24.99\\
                                           &$\rho K_1(1400)$               & 8.29              & 0.19\\
                                           &$\phi K_1(1270)$         & 2.45              & 0.06\\
                                           &$K^* h_1(1380)$                & 2.40              & 0.63\\
                                           &$K^* f_1(1420)$                & 2.69              & 1.20\\
                                           &$\omega K_1(1270)$             & 1.16              & 7.95\\
                                           &$\omega K_1(1400)$             & 2.74              & 0.03\\
                                           &$K^* h_1(1170)$                & 2.01              & 2.92\\
                                           &$K^* f_1(1285)$                & 1.21              & 1.73\\
  \hline
  $0^-\rightarrow 1^-2^+ $                 &$K^* a_2(1320)$                & 8.40              & 1.71\\
                                           &$\rho K^*_2(1430) $             & 8.92              & 0.44\\
                                           &$\omega K^*_2(1430) $           & 2.92              & 0.23\\
                                           &$K^* f^\prime_2(1525)$          & 0.00006           & 0.000008\\
                                           &$K^* f_2(1270)$                & 3.05              & 0.07\\
  \hline
 \multicolumn{2}{c|}{Total width}        &  163.38    & 224.98\\
\hline\hline
\end{tabular}
\end{center}
\end{table}

\begin{figure}[htb]
\centering
\includegraphics[scale=0.5]{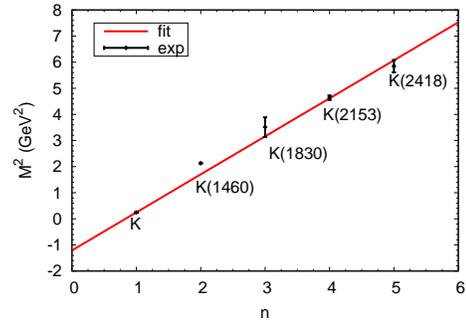}
\caption{The $K$-trajectory with $M^2_n=M^2_0+(n-1)\mu^2$. $M^2_0=0.24669\pm 0.00002 \mbox{GeV}^2$, $\mu^2=1.456\pm 0.026\mbox{GeV}^2$, $\chi^2/\mbox{d.o.f}=1.897/(4-2)$. In our fit, the we don't use the data of $K(1460)$ since the $K(1460)$ mass error is not given experimentally. The masses of the
$K$ and $K(1830)$ are taken from Ref.~\cite{PDG2016}. The masses of the $K(2153)$ and $K(2418)$ are taken to be $2153\pm 20$ MeV and $2418\pm 49$ MeV, respectively. The $K(1460)$ mass is taken to be 1460 MeV~\cite{Daum:1981hb}.}
\label{fig:regge_kaon}
\end{figure}

As mentioned in Sec.~\ref{sec:eta2320}, with the $\pi(2360)$, $\eta(2320)$, and $X(2500)$ as the members of $5^1S_0$ meson nonet, from Eq.~(\ref{msch}), the $K(5^1S_0)$ mass is predicted to be about 2418 MeV as shown in Eqs.~(\ref{eq: Kmass1}) and (\ref{eq:Kmass2}). At present, no candidate for the $I(J^P)=1/2(0^{-})$ state around 2418 MeV is reported
experimentally. It is noted that with our estimated masses for the $K(4^1S_0)$ and $K(5^1S_0)$,  $M_{K(4^1S_0)}=2153\pm 20$ MeV~\cite{Li:2008et} and $M_{K(5^1S_0)}=2418\pm 49$ MeV, the $K$, $K(1460)$, $K(1830)$, $K(2153)$, and $K(2418)$ approximately populate a trajectory as shown in Fig.~\ref{fig:regge_kaon}, which indicates that the $K(2153)$ and $K(2418)$ could be the good candidates for the $4^1S_0$ and $5^1S_0$ kaons, respectively.

The decay widths of the $K(2418)$ as the $5^1S_0$ kaon are listed in Table~\ref{tab:2420}. The total width of the $K(5^1S_0)$ is predicated to be about 163 MeV in case A or 225~MeV in case B. This could be of use in looking for the
candidate for the $5^1S_0$ kaon experimentally.

\section{SUMMARY AND CONCLUSION}
\label{sec:summary}

In this work, we have discussed the possible members of the $5^1S_0$ meson nonet by analysing the masses and calculating the strong decay widths in the $^3P_0$ model with the SHO and RQM meson space wave functions. Both the mass and width for the $\pi(2360)$ are consistent with the quark model expectations for the $\pi(5^1S_0)$. In the presence of the $X(2500)$ as the $5^1S_0$ isoscalar state, the possibility of the $\eta(2320)$ and $X(2370)$ as the isoscalar partner of the $X(2500)$ is discussed. The $X(2370)$ width can not be reproduced for any value of the mixing angle $\phi$, thus, the assignment of the $X(2370)$ as the $5^1S_0$ isoscalar state is not favored by its width. Both the $\eta(2320)$ and $X(2500)$ widths can be reproduced with $-0.5\leq \phi \leq 0.45$ radians for the SHO wave functions or $-0.69\leq \phi \leq 0.59$ for the RQM wave functions. The assignment of the $\pi(2360)$, $\eta(2320)$, and $X(2500)$ as the members of the $5^1S_0$ nonet not only gives $\phi=-0.1$ radians, which naturally accounts for the $\eta(2320)$ and $X(2500)$ widths, but also shows that the $5^1S_0$ kaon has a mass of about 2418 MeV. The $K$, $K(1460)$, $K(1830)$, $K(2153)$, and $K(2418)$ approximately populate a common trajectory. The $K(2418)$ is predicted to have a width of about 163 MeV for SHO wave functions or 225 MeV for the RQM wave functions. We tend to conclude that the $\pi(2360)$, $\eta(2320)$, $X(2500)$, together with the unobserved $K(2418)$, construct the $5^1S_0$ meson nonet.

Our numerical results show that the $^3P_0$ model predictions depend on the choice of meson space wave functions. It is essential to treat the wave functions accurately in the $^3P_0$ model calculations. The difference between the predictions in SHO case and those in RQM case provides a chance to distinguish among different meson space wave functions. To conclude which type of wave function is preferable, the further experimental study on the decays of $\pi(2360)$, $\eta(2320)$, and $X(2500)$ is needed. Also, in our calculations, all the states are assumed to be $q\bar{q}$. It is noted that some resonances such as $h_1(1170)$, $h_1(1380)$, $f_1(1285)$, $b_1(1235)$, $a_1(1260)$, and $K_1(1270)$, can also be explained as the dynamically generated resonances~\cite{Roca:2005nm,Geng:2006yb,GarciaRecio:2010ki}, which means they might have large hadron-molecular components in their wave functions. If so, both the SHO and RQM wave functions derived from the simple $q\bar{q}$ picture, would be not appropriate and could lead to the big discrepancies between the experiments and the $^3P_0$ model predictions. To test this point, the further experimental information about the partial widths is also needed.

\bigskip
\noindent
\begin{center}
{\bf ACKNOWLEDGEMENTS}\\
\end{center}
We would like to thank Dr. Qi-Fang L\"{u} for valuable discussions. This work is partly supported by the National Natural Science Foundation of China under Grant Nos. 11505158, 11605158, the China Postdoctoral Science Foundation under Grant No.2015M582197, the Postdoctoral Research Sponsorship in Henan Province under Grant No.2015023, and the Academic Improvement Project
of Zhengzhou University.


\begin{thebibliography}{99}


%\cite{Ablikim:2010au}
\bibitem{Ablikim:2010au}
 M.~Ablikim {\it et al.} [BESIII Collaboration],
 Confirmation of the $X(1835)$ and observation of the resonances $X(2120)$ and $X(2370)$ in $J/\psi\to \gamma \pi^+\pi^-\eta^\prime$,
 Phys.\ Rev.\ Lett.\  {\bf 106}, 072002 (2011).
%  doi:10.1103/PhysRevLett.106.072002
%  [arXiv:1012.3510 [hep-ex]].
  %%CITATION = doi:10.1103/PhysRevLett.106.072002;%%
  %100 citations counted in INSPIRE as of 10 Jan 2018


%\cite{PDG2016}
\bibitem{PDG2016}
  C.~Patrignani {\it et al.} [Particle Data Group],
  Review of Particle Physics,
  Chin.\ Phys.\ C {\bf 40}, no. 10, 100001 (2016).
%  doi:10.1088/1674-1137/40/10/100001
  %%CITATION = doi:10.1088/1674-1137/40/10/100001;%%
  %699 citations counted in INSPIRE as of 21 Apr 2017


\bibitem{Ablikim:2016hlu}
M.~Ablikim {\it et al.} [BESIII Collaboration],
Observation of pseudoscalar and tensor resonances in $J/\psi\to \gamma \phi \phi$,
Phys.\ Rev.\ D {\bf 93}, no. 11, 112011 (2016).
%doi:10.1103/PhysRevD.93.112011 ??[arXiv:1602.01523 [hep-ex]]. ??%%CITATION = doi:10.1103/PhysRevD.93.112011;%%   %14 citations counted in INSPIRE as of 03 Feb 2018



  %\cite{Li:2008mza}
\bibitem{Li:2008mza}
  D.~M.~Li and B.~Ma,
  $X(1835)$ and $\eta(1760)$ observed by the BES Collaboration,
  Phys.\ Rev.\ D {\bf 77}, 074004 (2008).
 % doi:10.1103/PhysRevD.77.074004
  %[arXiv:0801.4821 [hep-ph]].
  %%CITATION = doi:10.1103/PhysRevD.77.074004;%%
  %26 citations counted in INSPIRE as of 23 Mar 2016

%\cite{Yu:2011ta}
%\bibitem{Yu:2011ta}
%  J.~S.~Yu, Z.~F.~Sun, X.~Liu and Q.~Zhao,
%  Categorizing resonances X(1835), X(2120) and X(2370) in the pseudoscalar meson family,
%  Phys.\ Rev.\ D {\bf 83}, 114007 (2011).
 % doi:10.1103/PhysRevD.83.114007
  %[arXiv:1104.3064 [hep-ph]].
  %%CITATION = doi:10.1103/PhysRevD.83.114007;%%
  %17 citations counted in INSPIRE as of 06 Apr 2016

 %\cite{Li:2008et}
\bibitem{Li:2008et}
  D.~M.~Li and S.~Zhou,
  Towards the assignment for the $4^1S_0$ meson nonet,
  Phys.\ Rev.\ D {\bf 78}, 054013 (2008).
%  doi:10.1103/PhysRevD.78.054013
 % [arXiv:0805.3404 [hep-ph]].
  %%CITATION = doi:10.1103/PhysRevD.78.054013;%%
  %12 citations counted in INSPIRE as of 23 Mar 2016

%\cite{Pan:2016bac}
\bibitem{Pan:2016bac}
  T.~T.~Pan, Q.~F.~L\"{u} , E.~Wang and D.~M.~Li,
  Strong decays of the $X(2500)$ newly observed by the BESIII Collaboration,
  Phys.\ Rev.\ D {\bf 94}, no. 5, 054030 (2016).
%  doi:10.1103/PhysRevD.94.054030
%  [arXiv:1606.08635 [hep-ph]].
%%CITATION = doi:10.1103/PhysRevD.94.054030;%%

%\cite{Morningstar:1999rf}
\bibitem{Morningstar:1999rf}
C.~J.~Morningstar and M.~J.~Peardon,
The Glueball spectrum from an anisotropic lattice study,
Phys.\ Rev.\ D {\bf 60}, 034509 (1999).
%doi:10.1103/PhysRevD.60.034509 ??[hep-lat/9901004]. ??%%CITATION = doi:10.1103/PhysRevD.60.034509;%%   %799 citations counted in INSPIRE as of 08 Feb 2018

%\cite{Hart:2001fp}
 \bibitem{Hart:2001fp}
 A.~Hart {\it et al.} [UKQCD Collaboration],
 On the glueball spectrum in $O(a)$ improved lattice QCD,
 Phys.\ Rev.\ D {\bf 65}, 034502 (2002).
 % doi:10.1103/PhysRevD.65.034502 ??[hep-lat/0108022]. ??%%CITATION = doi:10.1103/PhysRevD.65.034502;%%   %76 citations counted in INSPIRE as of 08 Feb 2018

%\cite{Chen:2005mg}
 \bibitem{Chen:2005mg}
 Y.~Chen {\it et al.},
 Glueball spectrum and matrix elements on anisotropic lattices,
 Phys.\ Rev.\ D {\bf 73}, 014516 (2006).
 %doi:10.1103/PhysRevD.73.014516 ??[hep-lat/0510074]. ??%%CITATION = doi:10.1103/PhysRevD.73.014516;%%   %404 citations counted in INSPIRE as of 08 Feb 2018



\bibitem{Anisovich:2001pn}
 A.~V.~Anisovich, C.~A.~Baker, C.~J.~Batty, D.~V.~Bugg, V.~A.~Nikonov, A.~V.~Sarantsev, V.~V.~Sarantsev, and B.~S.~Zou,
 Partial wave analysis of $\bar{p}p$ annihilation channels in flight with $I = 1, C = +1$, Phys.\ Lett.\ B {\bf 517}, 261 (2001).

%\cite{Anisovich:2001pp}
\bibitem{Anisovich:2001pp}
  A.~V.~Anisovich, C.~A.~Baker, C.~J.~Batty, D.~V.~Bugg, V.~A.~Nikonov, A.~V.~Sarantsev, V.~V.~Sarantsev and B.~S.~Zou,
  A partial wave analysis of $\bar{p} p\to \eta \eta \pi^0$,
  Phys.\ Lett.\ B {\bf 517}, 273 (2001).
%  doi:10.1016/S0370-2693(01)01018-8
%  [arXiv:1109.6817 [hep-ex]].
  %%CITATION = doi:10.1016/S0370-2693(01)01018-8;%%
  %20 citations counted in INSPIRE as of 02 Mar 2017


\bibitem{Khruschov:2005zc}
V.~V.~Khruschov,
Mass spectra of excited meson states consisting of $u$-, $d$-quarks and antiquarks, arXiv:hep-ph/0504077.

 \bibitem{Ebert:2009ub}
 D.~Ebert, R.~N.~Faustov, and V.~O.~Galkin,
 Mass spectra and Regge trajectories of light mesons in the relativistic quark model, Phys.\ Rev.\ D {\bf 79}, 114029 (2009).



%\cite{Anisovich:2000kxa}
\bibitem{Anisovich:2000kxa}
  A.~V.~Anisovich, V.~V.~Anisovich and A.~V.~Sarantsev,
  Systematics of $q\bar{q}$ states in the $(n, M^2)$ and $(J, M^2)$ planes,
  Phys.\ Rev.\ D {\bf 62}, 051502 (2000).
%  doi:10.1103/PhysRevD.62.051502
%  [hep-ph/0003113].
  %%CITATION = doi:10.1103/PhysRevD.62.051502;%%
  %236 citations counted in INSPIRE as of 09 Jan 2018

%\cite{Anisovich:2001ig}
\bibitem{Anisovich:2001ig}
V.~V.~Anisovich,
Systematics of $q\bar{q}$ states, scalar mesons and glueball,
AIP Conf.\ Proc.\  {\bf 619}, 197 (2002).
%doi:10.1063/1.1482449 ??[hep-ph/0110326]. ??%%CITATION = doi:10.1063/1.1482449;%%   %30 citations counted in INSPIRE as of 03 Feb 2018

%\cite{Anisovich:2002us}
 \bibitem{Anisovich:2002us}
 V.~V.~Anisovich,
 Systematics of quark anti-quark states and scalar exotic mesons,
 Phys.\ Usp.\  {\bf 47}, 45 (2004) [Usp.\ Fiz.\ Nauk {\bf 47}, 49 (2004)].
 %?doi:10.1070/PU2004v047n01ABEH001333 ??[hep-ph/0208123]. ??%%CITATION = doi:10.1070/PU2004v047n01ABEH001333;%%   %69 citations counted in INSPIRE as of 03 Feb 2018

%\cite{Anisovich:2003tm}
\bibitem{Anisovich:2003tm}
V.~V.~Anisovich,
Systematics of quark anti-quark states: Where are the lightest glueballs?,
AIP Conf.\ Proc.\  {\bf 717}, 441 (2004).
%doi:10.1063/1.1799747 ??[hep-ph/0310165]. ??%%CITATION = doi:10.1063/1.1799747;%%   %21 citations counted in INSPIRE as of 03 Feb 2018

%\cite{Anisovich:2004vj}
\bibitem{Anisovich:2004vj}
V.~V.~Anisovich,
Systematization of tensor mesons and the determination of the $2^{++}$ glueball,
JETP Lett.\  {\bf 80}, 715 (2004) [Pisma Zh.\ Eksp.\ Teor.\ Fiz.\  {\bf 80}, 845 (2004)].
 %??doi:10.1134/1.1868792 ??[hep-ph/0412093]. ??%%CITATION = doi:10.1134/1.1868792;%%   %16 citations counted in INSPIRE as of 03 Feb 2018

%\cite{collins}
\bibitem{collins}
P. D. Collins, An introduction to Regge theory and high energy physics, Cambridge University  Press, 1977.

%\cite{Li:2007px}
\bibitem{Li:2007px}
D.~M.~Li, B.~Ma, and Y.~H.~Liu,
Understanding masses of $c\bar{s}$ states in Regge phenomenology,
Eur.\ Phys.\ J.\ C {\bf 51}, 359 (2007).
%doi:10.1140/epjc/s10052-007-0286-7 ??[hep-ph/0703278 [HEP-PH]]. ??%%CITATION = doi:10.1140/epjc/s10052-007-0286-7;%%   %28 citations counted in INSPIRE as of 06 Feb 2018


%\cite{Roberts:1992js}
\bibitem{Roberts:1992js}
    W.~Roberts and B.~Silvestre-Brac,
    General method of calculation of any hadronic decay in the $^3P_0$ triplet model,
    Few-Body Syst. 11, 171 (1992).
     %%CITATION = APASA,11,171;%%   %9 citations counted in INSPIRE as of 22 Dec 2017

%\cite{Blundell:1996as}
\bibitem{Blundell:1996as}
  H.~G.~Blundell,
  Meson properties in the quark model: A look at some outstanding problems,
  hep-ph/9608473.
  %%CITATION = HEP-PH/9608473;%%
  %25 citations counted in INSPIRE as of 23 Mar 2016


%\cite{Barnes:1996ff}
 \bibitem{Barnes:1996ff}
     T.~Barnes, F.~E.~Close, P.~R.~Page, and E.~S.~Swanson,
        Higher quarkonia,
 Phys.\ Rev.\ D {\bf 55}, 4157 (1997).
  %??doi:10.1103/PhysRevD.55.4157 ??[hep-ph/9609339]. ??%%CITATION = doi:10.1103/PhysRevD.55.4157;%%   %228 citations counted in INSPIRE as of 22 Dec 2017

%\cite{Barnes:2002mu}
\bibitem{Barnes:2002mu}
  T.~Barnes, N.~Black, and P.~R.~Page,
  Strong decays of strange quarkonia,
  Phys.\ Rev.\ D {\bf 68}, 054014 (2003).
%  doi:10.1103/PhysRevD.68.054014
%  [nucl-th/0208072].
  %%CITATION = doi:10.1103/PhysRevD.68.054014;%%
  %115 citations counted in INSPIRE as of 28 Sep 2017

%\cite{Close:2005se}
\bibitem{Close:2005se}
    F.~E.~Close and E.~S.~Swanson,
Dynamics and decay of heavy-light hadrons,
Phys.\ Rev.\ D {\bf 72}, 094004 (2005).
         %??doi:10.1103/PhysRevD.72.094004 ??[hep-ph/0505206]. ??%%CITATION = doi:10.1103/PhysRevD.72.094004;%%   %152 citations counted in INSPIRE as of 22 Dec 2017


%\cite{Barnes:2005pb}
\bibitem{Barnes:2005pb}
    T.~Barnes, S.~Godfrey, and E.~S.~Swanson,
       Higher charmonia,
        Phys.\ Rev.\ D {\bf 72}, 054026 (2005).
         %??doi:10.1103/PhysRevD.72.054026 ??[hep-ph/0505002]. ??%%CITATION = doi:10.1103/PhysRevD.72.054026;%%   %471 citations counted in INSPIRE as of 22 Dec 2017



%\cite{Zhang:2006yj}
\bibitem{Zhang:2006yj}
    B.~Zhang, X.~Liu, W.~Z.~Deng, and S.~L.~Zhu,
       $D_{sJ}(2860)$ and $D_{sJ}(2715)$,
        Eur.\ Phys.\ J.\ C {\bf 50}, 617 (2007).
         %??doi:10.1140/epjc/s10052-007-0221-y ??[hep-ph/0609013]. ??%%CITATION = doi:10.1140/epjc/s10052-007-0221-y;%%   %102 citations counted in INSPIRE as of 22 Dec 2017



%\cite{Ding:2007pc}
\bibitem{Ding:2007pc}
    G.~J.~Ding and M.~L.~Yan,
       $Y(2175)$: Distinguish Hybrid State from Higher Quarkonium,
        Phys.\ Lett.\ B {\bf 657}, 49 (2007).
        %??doi:10.1016/j.physletb.2007.10.020 ??[hep-ph/0701047]. ??%%CITATION = doi:10.1016/j.physletb.2007.10.020;%%   %60 citations counted in INSPIRE as of 22 Dec 2017

\bibitem{Li:2008we}
  D.~M.~Li and B.~Ma,
 The $\eta(2225)$ observed by the BES Collaboration,
  Phys.\ Rev.\ D {\bf 77}, 094021 (2008).
%  doi:10.1103/PhysRevD.77.094021
 % [arXiv:0803.0106 [hep-ph]].
  %%CITATION = doi:10.1103/PhysRevD.77.094021;%%
  %19 citations counted in INSPIRE as of 09 Jun 2017


%\cite{Li:2008xy}
\bibitem{Li:2008xy}
D.~M.~Li and S.~Zhou,
     Nature of the $\pi_2(1880)$
      Phys.\ Rev.\ D {\bf 79}, 014014 (2009).
%       ??doi:10.1103/PhysRevD.79.014014 ??[arXiv:0811.0918 [hep-ph]]. ??%%CITATION = doi:10.1103/PhysRevD.79.014014;%%   %16 citations counted in INSPIRE as of 22 Dec 2017

%\cite{Li:2009rka}
\bibitem{Li:2009rka}
  D.~M.~Li and E.~Wang,
  Canonical interpretation of the $\eta_2(1870)$,
  Eur.\ Phys.\ J.\ C {\bf 63},297 (2009).
%  doi:10.1140/epjc/s10052-009-1106-z
 % [arXiv:0904.1252 [hep-ph]].
  %%CITATION = doi:10.1140/epjc/s10052-009-1106-z;%%
  %5 citations counted in INSPIRE as of 24 Mar 2016
%\cite{Li:2008we}


%\cite{Li:2009qu}
 \bibitem{Li:2009qu}
 D.~M.~Li and B.~Ma,
 Implication of BaBar's new data on the $D_{s1}(2710)$ and $D_{sJ}(2860)$,
Phys.\ Rev.\ D {\bf 81}, 014021 (2010).
 %??doi:10.1103/PhysRevD.81.014021 ??[arXiv:0911.2906 [hep-ph]]. ??%%CITATION = doi:10.1103/PhysRevD.81.014021;%%   %36 citations counted in INSPIRE as of 22 Dec 2017

%\cite{Li:2010vx}
\bibitem{Li:2010vx}
D.~M.~Li, P.~F.~Ji, and B.~Ma,
  The newly observed open-charm states in quark model,
   Eur.\ Phys.\ J.\ C {\bf 71}, 1582 (2011).
    %??doi:10.1140/epjc/s10052-011-1582-9 ??[arXiv:1011.1548 [hep-ph]]. ??%%CITATION = doi:10.1140/epjc/s10052-011-1582-9;%%   %45 citations counted in INSPIRE as of 22 Dec 2017

%\cite{Lu:2014zua}
\bibitem{Lu:2014zua}
    Q.~F.~L\"{u} and D.~M.~Li,
Understanding the charmed states recently observed by the LHCb and BaBar Collaborations in the quark model,
Phys.\ Rev.\ D {\bf 90}, 054024 (2014).
     %??doi:10.1103/PhysRevD.90.054024 ??[arXiv:1407.3092 [hep-ph]]. ??%%CITATION = doi:10.1103/PhysRevD.90.054024;%%   %18 citations counted in INSPIRE as of 22 Dec 2017


%\cite{Lu:2016bbk}
 \bibitem{Lu:2016bbk}
   Q.~F.~L\"{u}, T.~T.~Pan, Y.~Y.~Wang, E.~Wang, and D.~M.~Li,
 Excited bottom and bottom-strange mesons in the quark model,
Phys.\ Rev.\ D {\bf 94},074012 (2016).
      %??doi:10.1103/PhysRevD.94.074012 ??[arXiv:1607.02812 [hep-ph]]. ??%%CITATION = doi:10.1103/PhysRevD.94.074012;%%   %9 citations counted in INSPIRE as of 22 Dec 2017

%\cite{Wang:2017pxm}
 \bibitem{Wang:2017pxm}
 G.~Y.~Wang, S.~C.~Xue, G.~N.~Li, E.~Wang, and D.~M.~Li,
 Strong decays of the higher isovector scalar mesons,
 Phys.\ Rev.\ D {\bf 97}, no. 3, 034030 (2018).
 %doi:10.1103/PhysRevD.97.034030 ??[arXiv:1712.10180 [hep-ph]]. ??%%CITATION = doi:10.1103/PhysRevD.97.034030;%%


\bibitem{Hayne:1981zy}
  C.~Hayne and N.~Isgur,
  Beyond the Wave Function at the Origin: Some Momentum Dependent Effects in the Nonrelativistic Quark Model,
  Phys.\ Rev.\ D {\bf 25}, 1944 (1982).
 % doi:10.1103/PhysRevD.25.1944
  %%CITATION = doi:10.1103/PhysRevD.25.1944;%%
  %157 citations counted in INSPIRE as of 24 Mar 2016

\bibitem{Jacob:1959at}
 M.~Jacob and G.~C.~Wick,
 On the general theory of collisions for particles with spin,
 Annals Phys.\ {\bf 7}, 404 (1959).


%\cite{Micu:1968mk}
\bibitem{Micu:1968mk}
  L.~Micu,
  Decay rates of meson resonances in a quark model,
  Nucl.\ Phys.\ B {\bf 10}, 521 (1969).
  %doi:10.1016/0550-3213(69)90039-X
  %%CITATION = doi:10.1016/0550-3213(69)90039-X;%%
  %317 citations counted in INSPIRE as of 24 Mar 2016

%\cite{Godfrey:1985xj}
\bibitem{Godfrey:1985xj}
  S.~Godfrey and N.~Isgur,
  Mesons in a Relativized Quark Model with Chromodynamics,
  Phys.\ Rev.\ D {\bf 32}, 189 (1985).
%  doi:10.1103/PhysRevD.32.189
  %%CITATION = doi:10.1103/PhysRevD.32.189;%%
  %2332 citations counted in INSPIRE as of 27 Apr 2017


%\cite{Ackleh:1996yt}
\bibitem{Ackleh:1996yt}
  E.~S.~Ackleh, T.~Barnes and E.~S.~Swanson,
  On the mechanism of open flavor strong decays,
  Phys.\ Rev.\ D {\bf 54}, 6811 (1996).
 % doi:10.1103/PhysRevD.54.6811
  %[hep-ph/9604355].
  %%CITATION = doi:10.1103/PhysRevD.54.6811;%%
  %200 citations counted in INSPIRE as of 24 Mar 2016



%\cite{Wang:2017iai}
\bibitem{Wang:2017iai}
  L.~M.~Wang, S.~Q.~Luo, Z.~F.~Sun and X.~Liu,
  Constructing new pseudoscalar meson nonets with the observed $X(2100)$, $X(2500)$, and $\eta(2225)$,
  Phys.\ Rev.\ D {\bf 96}, no. 3, 034013 (2017).
%  doi:10.1103/PhysRevD.96.034013
%  [arXiv:1705.00549 [hep-ph]].
  %%CITATION = doi:10.1103/PhysRevD.96.034013;%%

%\cite{Blundell:1995ev}
\bibitem{Blundell:1995ev}
H.~G.~Blundell and S.~Godfrey,
The $\xi(2220)$ revisited: Strong decays of the $1^3F_2$ and $1^3F_4$ $s\bar{s}$ mesons,
Phys.\ Rev.\ D {\bf 53}, 3700 (1996).
%doi:10.1103/PhysRevD.53.3700 ??[hep-ph/9508264]. ??%%CITATION = doi:10.1103/PhysRevD.53.3700;%%   %104 citations counted in INSPIRE as of 06 Feb 2018

%\cite{Kokoski:1985is}
\bibitem{Kokoski:1985is}
R.~Kokoski and N.~Isgur,
Meson Decays by Flux Tube Breaking,
Phys.\ Rev.\ D {\bf 35}, 907 (1987).
%??doi:10.1103/PhysRevD.35.907 ??%%CITATION = doi:10.1103/PhysRevD.35.907;%%   %434 citations counted in INSPIRE as of 08 May 2018

%\cite{Lu:2016mbb}
\bibitem{Lu:2016mbb}
Y.~Lu, M.~N.~Anwar, and B.~S.~Zou,
Coupled-Channel Effects for the Bottomonium with Realistic Wave Functions,
Phys.\ Rev.\ D {\bf 94}, no. 3, 034021 (2016).
%doi:10.1103/PhysRevD.94.034021 ??[arXiv:1606.06927 [hep-ph]]. ??%%CITATION = doi:10.1103/PhysRevD.94.034021;%%   %10 citations counted in INSPIRE as of 09 May 2018


%\cite{Anisovich:2000ix}
\bibitem{Anisovich:2000ix}
  A.~V.~Anisovich, C.~A.~Baker, C.~J.~Batty, D.~V.~Bugg, V.~A.~Nikonov, A.~V.~Sarantsev, V.~V.~Sarantsev, and B.~S.~Zou,
  A study of $\bar{p}p \to \eta \eta \eta$ for masses $1960$-$2410$ MeV/c$^2$,
  Phys.\ Lett.\ B {\bf 496}, 145 (2000).
%  doi:10.1016/S0370-2693(00)01301-0
  %%CITATION = doi:10.1016/S0370-2693(00)01301-0;%%
  %13 citations counted in INSPIRE as of 10 Jan 2018


%\cite{Li:2001qg}
 \bibitem{Li:2001qg}
 D.~M.~Li, H.~Yu, and Q.~X.~Shen,
 Effects of flavor dependent $\bar{q}q$ annihilation on the mixing angle of the isoscalar octet
 singlet and Schwinger's nonet mass formula,
 Chin.\ Phys.\ Lett.\  {\bf 18}, 184 (2001).
  %doi:10.1088/0256-307X/18/2/310 ??[hep-ph/0104102]. ??%%CITATION = doi:10.1088/0256-307X/18/2/310;%%   %8 citations counted in INSPIRE as of 07 Feb 2018

%\cite{Schwinger:1964zza}
 \bibitem{Schwinger:1964zza}
 J.~Schwinger,
 A Ninth Baryon?,
 Phys.\ Rev.\ Lett.\  {\bf 12}, 237 (1964).
 % doi:10.1103/PhysRevLett.12.237 ??%%CITATION = doi:10.1103/PhysRevLett.12.237;%%   %98 citations counted in INSPIRE as of 08 Feb 2018



%\cite{Yu:2011ta}
\bibitem{Yu:2011ta}
J.~S.~Yu, Z.~F.~Sun, X.~Liu, and Q.~Zhao,
Categorizing resonances $X(1835)$, $X(2120)$ and $X(2370)$ in the pseudoscalar meson family,
Phys.\ Rev.\ D {\bf 83}, 114007 (2011).
%doi:10.1103/PhysRevD.83.114007 ??[arXiv:1104.3064 [hep-ph]]. ??%%CITATION = doi:10.1103/PhysRevD.83.114007;%%   %27 citations counted in INSPIRE as of 08 Feb 2018



%\cite{Liu:2010tr}
\bibitem{Liu:2010tr}
J.~F.~Liu {\it et al.} [BES Collaboration],
 $X(1835)$ and the New Resonances $X(2120)$ and $X(2370)$ Observed by the BES Collaboration,
 Phys.\ Rev.\ D {\bf 82}, 074026 (2010).
 %doi:10.1103/PhysRevD.82.074026 ??[arXiv:1008.0246 [hep-ph]]. ??%%CITATION = doi:10.1103/PhysRevD.82.074026;%%   %13 citations counted in INSPIRE as of 08 Feb 2018

%\cite{Chen:2011kp}
 \bibitem{Chen:2011kp}
 S.~Chen and J.~Ping,
 Radial excitation states of $\eta$ and $\eta^\prime$ in the chiral quark model,
 Chin.\ Phys.\ C {\bf 36}, 681 (2012).
 %doi:10.1088/1674-1137/36/8/001 ??[arXiv:1111.3533 [hep-ph]]. ??%%CITATION = doi:10.1088/1674-1137/36/8/001;%%   %7 citations counted in INSPIRE as of 08 Feb 2018

%\cite{Wang:2010vz}
 \bibitem{Wang:2010vz}
 Z.~G.~Wang,
 Analysis of the $X(1835)$ and related baryonium states with Bethe-Salpeter equation,
 Eur.\ Phys.\ J.\ A {\bf 47}, 71 (2011).
 %doi:10.1140/epja/i2011-11071-x ??[arXiv:1012.3325 [hep-ph]]. ??%%CITATION = doi:10.1140/epja/i2011-11071-x;%%   %4 citations counted in INSPIRE as of 08 Feb 2018





%\cite{Daum:1981hb}
\bibitem{Daum:1981hb}
C.~Daum {\it et al.} [ACCMOR Collaboration],
Diffractive Production of Strange Mesons at 63-{GeV},
Nucl.\ Phys.\ B {\bf 187}, 1 (1981).
%doi:10.1016/0550-3213(81)90114-0 ??%%CITATION = doi:10.1016/0550-3213(81)90114-0;%%   %137 citations counted in INSPIRE as of 09 May 2018

%\cite{Roca:2005nm}
\bibitem{Roca:2005nm}
L.~Roca, E.~Oset, and J.~Singh,
Low lying axial-vector mesons as dynamically generated resonances,
Phys.\ Rev.\ D {\bf 72}, 014002 (2005).
 %?doi:10.1103/PhysRevD.72.014002 ??[hep-ph/0503273]. ??%%CITATION = doi:10.1103/PhysRevD.72.014002;%%   %161 citations counted in INSPIRE as of 09 May 2018

%\cite{Geng:2006yb}
 \bibitem{Geng:2006yb}
 L.~S.~Geng, E.~Oset, L.~Roca, and J.~A.~Oller,
 Clues for the existence of two $K_1(1270)$ resonances,
 Phys.\ Rev.\ D {\bf 75}, 014017 (2007).
  %??doi:10.1103/PhysRevD.75.014017 ??[hep-ph/0610217]. ??%%CITATION = doi:10.1103/PhysRevD.75.014017;%%   %38 citations counted in INSPIRE as of 09 May 2018

%\cite{GarciaRecio:2010ki}
 \bibitem{GarciaRecio:2010ki}
 C.~Garcia-Recio, L.~S.~Geng, J.~Nieves, and L.~L.~Salcedo,
 Low-lying even parity meson resonances and spin-flavor symmetry,
 Phys.\ Rev.\ D {\bf 83}, 016007 (2011).
 %doi:10.1103/PhysRevD.83.016007 ??[arXiv:1005.0956 [hep-ph]]. ??%%CITATION = doi:10.1103/PhysRevD.83.016007;%%   %34 citations counted in INSPIRE as of 09 May 2018





\end{thebibliography}
\end{document}